\documentclass[%
 reprint,
 nofootinbib,
 amsmath,amssymb,
 aps,
 prx,
]{revtex4-2}

\usepackage[hidelinks]{hyperref}
\usepackage{graphicx}
\usepackage{dcolumn}
\usepackage{bm}
\usepackage{booktabs}
\usepackage{amsmath,amssymb,amsbsy,amstext, amsthm, amsfonts}
\usepackage{color}
\usepackage{slashed}
\usepackage[table,dvipsnames]{xcolor}
\usepackage{dsfont}
\usepackage{verbatim}
\usepackage[utf8]{inputenc} 
\usepackage{ragged2e}
\usepackage{mathtools}
\usepackage{xfrac}
\usepackage{blkarray}

\usepackage{tikz}
\usetikzlibrary{shapes.misc}
\usepackage[compat=1.1.0]{tikz-feynman}

\newcommand{\half}{\frac{1}{2}}

\def\Vhrulefill{\leavevmode\leaders\hrule height 0.7ex depth \dimexpr0.4pt-0.7ex\hfill\kern0pt}

\newcommand\extrafootertext[1]{%
    \bgroup
    \renewcommand\thefootnote{\fnsymbol{footnote}}%
    \renewcommand\thempfootnote{\fnsymbol{mpfootnote}}%
    \footnotetext[0]{#1}%
    \egroup
}

\begin{document}

\title{Quark-Lepton Color-Flavor Unification}

\author{Antonio Delgado}
\email{adelgad2@nd.edu}
\affiliation{Department of Physics and Astronomy, University of Notre Dame}
\author{Seth Koren}
\email{skoren@nd.edu}
\affiliation{Department of Physics and Astronomy, University of Notre Dame}

\date{\today}

\begin{abstract}
We present an $SU(12) \times SU(2)_L \times U(1)_R$ model unifying $SU(9)$ quark color-flavor with $SU(3)$ lepton flavor as a flavorful alternative to conventional theories of unification. 
We begin in the ultraviolet with a single yukawa shared by the unified up-type quarks and neutrinos, and no further new fermions. 
We show that gauged quark color-flavor and lepton flavor instantons dynamically generate the bottom and tau yukawas, which implements a massless quark solution to the strong CP problem and sets up a flavored type-I seesaw mechanism. Only two new scalar irreps are needed for the symmetry-breaking steps, which include quark color-flavor deconstruction and then infrared reunification, and the Standard Model gauge group in this theory emerges as \[G_{\rm SM} = \frac{SU(3)_C \times SU(2)_L \times U(1)_Y \times \mathbb{Z}^X_{18}}{\mathbb{Z}_{3} \times \Gamma \times \mathbb{Z}_3},\] where $\Gamma \in \lbrace 1, \mathbb{Z}_2 \rbrace$ is the electroweak global structure and there is a discrete gauge symmetry $X = B - 3(L_i + L_j - L_k)$ which brings additional $\mathbb{Z}_3$ global structure to the SM. This gauge symmetry acts as a flavorful upgrade of the $\mathbb{Z}^{B+L}_{18}$ anomaly-free global symmetry of the SM and stabilizes the proton absolutely. Non-invertible chiral symmetry-breaking is crucial to our model, and we discuss the rich spectrum of emergent generalized symmetries and topological defects appearing at various stages. In the infrared, the novel shared quotient between continuous and discrete groups links the one-form and two-form global symmetries of the Standard Model. 
\end{abstract}

\maketitle
\tableofcontents

\section{Unification} 

A major lesson of the past couple hundred years of scientific inquiry has been the rampant success of reductionism: As we have probed shorter and shorter distances, we have found simpler and simpler descriptions of the world and its remarkable emergent phenomena which require fewer degrees of freedom and fewer free parameters. With such large Bayesian update factors for the proposition of simpler small-distance theories, there is great \textit{circumstantial} evidence toward \textit{further} unification and an even simpler theory at distances far below the Standard Model phase.

However, since the proposal to unify the gauge forces of Georgi-Glashow in 1974 \cite{Georgi:1974sy}, particle physicists have sadly seen no discoveries of the concrete BSM microphysics involved in unification models.\footnote{Nor have we seen direct evidence of the proposals to unify gravity and SM forces through extra dimensions going back to N\"{o}rdstrom in 1914 \cite{Nordstrom:1914ejq}, followed later by Kaluza in 1921 \cite{Kaluza:1921tu} and Klein in 1926 \cite{Klein:1926tv}.} 
A general concern for theories that unify the SM gauge symmetries into a simple group is the lack of observed proton decay, despite exquisitely sensitive searches occurring over decades \cite{Super-Kamiokande:2016exg}. Of course it is possible the lesson is merely that unification happens at a higher scale than expected, but whenever there is conventional wisdom which is under stress from empirical observations, it is interesting to think contrarily and in particular to consider maximalist interpretations of the data. 

Maybe unification, despite the tremendous appeal, is just not realized in our universe: It turns out that if unification is \textit{not} correct, we could definitively learn this at low energies by discovering fractionally-charged particles (FCPs) \cite{Koren:2024xof,Koren:2025utp}, whose very existence is incompatible with some or all minimal GUTs, depending on the charge. 
The possibility of learning such stark, far-ultraviolet information from simple infrared measurements is enabled here because the existence of fractionally-charged particles probes a \textit{topological} aspect of the field theory, so does not depend on energy scale at all. Further contemplation on how to identify and make use of such \textit{rigid} QFT data contained in our theories of particle physics is warranted, and indeed we will make further use of topological data below.
Towards ruling out GUTs, more work by both phenomenologists and experimentalists is needed to develop a robust search program for these signatures at colliders.\footnote{See also \cite{Alonso:2024pmq,Li:2024nuo} on other aspects of FCPs.  We can learn similar information by discovering an axion and making measurements sensitive to its anomaly coefficients \cite{Cordova:2023her,Reece:2023iqn,Choi:2023pdp,Agrawal:2022lsp,Agrawal:2024ejr,Agrawal:2025rbr,Benabou:2025kgx,Reig:2025dqb} though to our knowledge the full map of what set of measurements would be required to rule out which global structures has not been described.} 

On the other hand, discarding unification entirely based on the non-observation of proton decay is perhaps too rash, and we ought also examine our theoretical biases regarding the link between unification and proton decay. Rather than giving up the prospect of a simpler theory at smaller distances, maybe there is a different approach to unification that leads to different conclusions about proton stability. After all, the proton of the Standard Model is \textit{stable} due to a zero-form discrete global symmetry that results non-trivially from the fact we have multiple generations of fermions \cite{Koren:2022bam,Wang:2022eag}. Here we follow to the maximum this hint that the multiplicity of generations is deeply tied to the fundamental behavior of the Standard Model particles, and find that non-trivially incorporating gauged flavor into our unification paradigm can absolutely stabilize the proton even with quarks and leptons unified in the ultraviolet.  

It is well known that models of unification where the UV group is non-simple need not introduce proton decay, with Pati-Salam \cite{Pati:1974yy} and trinification \cite{de1984trinification} the paradigmatic and minimal examples, although pursuing this requires care regarding which scalar representations are included, and can complicate the model building. However these models are often seen as unsatisfying since gauge unification is `not complete'. Is there a version of these that offers comparable virtues to well-loved $SU(5)$ \cite{Georgi:1974sy} and $SO(10)$ \cite{Fritzsch:1974nn}?

There are various ways in which theories of unification may simplify the ultraviolet description. Historically, the near-unification of SM gauge couplings has been important theoretical evidence \cite{Georgi:1974yf}, though more-precise measurements later revealed that the couplings do not unify with the SM matter \cite{Amaldi:1991cn,Giunti:1991ta,Langacker:1991an}. A conceptually separate aspect of unified theories is matter unification, and particularly the unification of the fermions of the SM. In the familiar `vertical' GUTs which focus on gauge coupling unification, the fermions unify into no less than three multiplets, as the generation structure merely goes along for the ride. Is there a way to achieve complete unification of the SM fermions? 

Attempts to add flavor to this picture include Georgi's program to enlarge $SU(5)$ to $SU(N>5)$ and find the generations from different representations to explain the breaking of flavor symmetry \cite{Georgi:1979md}. Berezhiani instead proposed vertical $\times$ horizontal theories like $SO(10) \times \mathcal{H}$ \cite{Berezhiani:1985in,Berezhiani:1993ub} which does unify the SM fermions into one multiplet. But either of these theories requires many new fermions, including perhaps new chiral states, and the proton anyway remains unstable. Furthermore, gauging flavor in this way requires having a shared flavor symmetry between the quarks and the leptons, as you might think is necessary if quarks and leptons are unified.

The `floccinaucinihilipilification' paper \cite{Allanach:2021bfe} of Allanach, Gripaios, and Tooby-Smith, betraying its name, uncovered a small number of ways in which gauged flavor can be intertwined nontrivially with the gauge structure of the SM. Davighi and Tooby-Smith constructed a fantastic theory of $SU(4)_C \times Sp(6)_L \times Sp(6)_R$ electroweak flavor unification \cite{Davighi:2022fer} (which had originally been noticed in 1984 by Kuo and Nakagawa \cite{Kuo:1984md, Kuo:1984gz} but without much exploration of the theory). This theory forbids perturbative proton decay but the proton inevitably decays through $Sp(6)$ instantons \cite{Koren:2026a}. The other sort of non-trivially intertwined gauge-flavor unification pointed out in that work is $SU(12) \times SU(2)_L \times SU(2)_R$ quark-lepton color-flavor unification. Note importantly that there exist versions of each of these theories which include a semi-direct product with a $\mathbb{Z}_2$ factor that charge-conjugates the color group while switching L and R.\footnote{In the context of Pati-Salam this is sometimes referred to as `D-parity', and the correct discrete structure (semi-direct product, \emph{not} direct) has to our knowledge been completely ignored in the literature.} In such a theory \textit{all} the SM fermions have truly been unified into a single irreducible representation, with no extras aside from the right-handed neutrinos. 

It is color-flavor unification that will be the topic of this work. Remarkably, despite manifestly unifying the leptons and the quarks, this will allow separate gauged flavor symmetries for quarks and leptons at intermediate scales. Not only does this provide a broad motivation for why quark and lepton flavor physics is so different, but it also allows remarkable nonperturbative effects from the gauged flavor instantons \cite{Cordova:2022fhg,Cordova:2024ypu,Delgado:2024pcv,Choi:2026oqz} which work specifically because the quarks and leptons are separated---and also specifically because we have $N_g = 3$ generations, making this one of few active inroads toward the generation puzzle. 

We have previously explored these sectors individually, having understood their effects from a bottom-up strategy using non-invertible symmetries which pointed us to interesting physics in these ultraviolet theories of unification. In those works, the neutrino yukawas are generated from the charged lepton yukawas and an instanton of $SU(3)_\ell$ lepton flavor unification, and the down-type yukawas are generated from the up-type yukawas and an instanton of $SU(9)$ quark color-flavor unification. Here we further unify these two separate uses of non-invertible symmetry in the quark and lepton sectors. However, there is an obstacle to directly cutting and pasting the previous uses together: The heaviest quark and lepton yukawas are reversed with respect to each other! That is, the `up-type' lepton is the \textit{neutrino}, which has the smaller of the two lepton masses, whereas the largest quark yukawa is the top, so the two mechanisms do not immediately unify. 
Here we will propose one model of notable dynamics coming from quark-lepton color-flavor unification with the expectation that this is one of many interesting phenomenologies that could be borne out as this ultraviolet theory flows to the infrared Standard Model. 
The model we will pursue in this work will preserve the solution to the strong CP problem while still using non-invertible symmetry breaking to interesting effect in the lepton sector in conjunction with a flavored type-I seesaw that ultimately leads to Majorana neutrino masses. 

\begin{figure}
    \centering
    \includegraphics[width=1\linewidth]{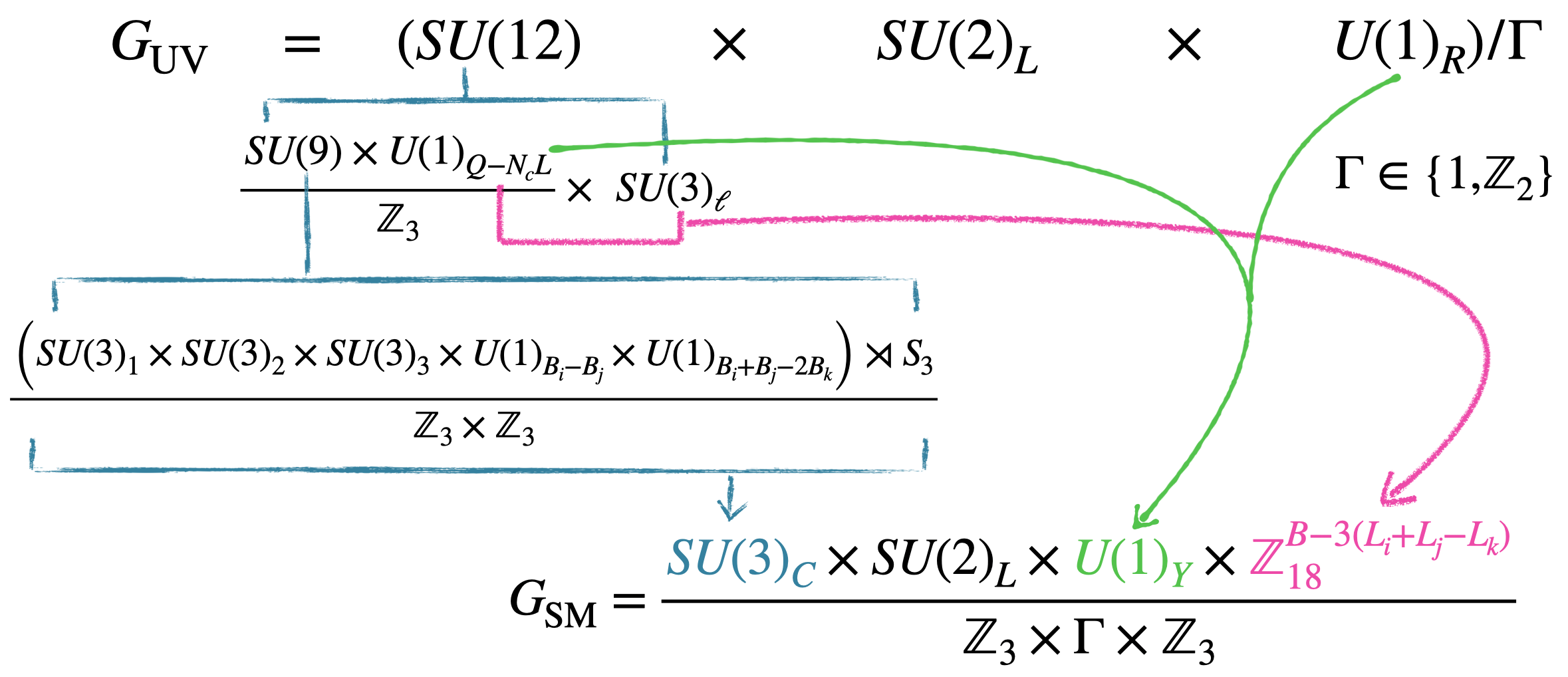}
    \caption{The breaking chain from our ultraviolet theory down to the Standard Model. Some of these steps may happen simultaneously; or contrarily there may be additional intermediate steps.}
    \label{fig:BreakingChart}
\end{figure}

Our main interest will be in starting with solely one yukawa and generating the various masses for the third-generation fermions through interplay between tree-level effects, one-loop effects, and nonperturbative effects. 
The penalty we pay with this approach is that quark-lepton unification will take place at very high energy scales, for reasons unrelated to the stability of the proton. After starting with the same yukawa for the top quark and the neutrino, we will generate the bottom quark and tau yukawas from gauged flavor instantons when those gauge symmetries are broken. But the breaking of this lepton flavor symmetry coincides with the generation of Majorana neutrino masses, which for a successful seesaw mechanism must occur at the usual high scale, necessitating very high energy dynamics. See Figure~\ref{fig:BreakingChart} for a schematic breakdown of the various symmetry-breaking steps from the ultraviolet group down to the SM phase.

As we flow down in energies and break the flavor symmetries, we will have enough freedom to generate the full structure of the masses and the CKM and PMNS matrices, but producing a beautiful model of flavor will be left for future work. As is standard in GUT model-building, in this first pass we will not attempt to write down a full scalar potential, but rather assume the appropriate scalars condense at the energies we want them to.

Nevertheless we will \textit{not} land on the familiar SM gauge group in the far IR---our Standard Model will have an additional discrete gauge theory! The full group will be 
\begin{equation}
    G_{\rm SM} = \frac{SU(3)_C \times SU(2)_L \times U(1)_Y \times \mathbb{Z}^X_{18}}{\mathbb{Z}_3 \times \Gamma \times \mathbb{Z}_3 } ,
\end{equation}
where $X = B - 3(L_i + L_j - L_k)$ will nonperturbatively protect the proton from decay and $\Gamma \in \lbrace 1, \mathbb{Z}_2 \rbrace$ is the electroweak global structure.\footnote{The $X$ quantum number is really best understood as $Q - 9(L_i + L_j - L_k) = 3[B - 3(L_i + L_j - L_k)]$, where $Q$ is quark number under which the quarks have unit charge. We've used `$B$' above because it's the familiar convention, and because two symmetries which differ only by normalization are the same symmetry, so it is not technically incorrect. But it is far less confusing to consider topological effects in the normalization where all particles have integer charge, so that a rotation by $2\pi$ is the identity and all anomaly coefficients and indices are integers, and in the rest of this work we will use unit normalizations. (That is, except for Section \ref{sec:matching} when we will switch to conventional normalization so that our results can be compared clearly to others.)} This discrete symmetry factor was essentially identified by Babu, Gogoladze, and Wang \cite{Babu:2003qh} long ago, and shown to appear from this theory in \cite{Davighi:2022qgb}. The additional global structure appearing between the continuous SM groups and a gauged discrete symmetry is novel here. This version of the Standard Model has the same continuous degrees of freedom as normal, and local-in-field-space effects are all the same as normal, but the spectrum of topological degrees of freedom differs and their coupling to the SM has important phenomenological implications.

In Section~\ref{sec:NonInvSym} we will review the logic of Non-invertible Naturalness, and the non-invertible symmetries appearing in this work. In Section~\ref{sec:uvSU12} we will give the ultraviolet theory of quark-lepton color-flavor unification. In Section~\ref{sec:breakingSU12} we will discuss the spontaneous breaking which separates the quarks and leptons and generates the non-invertible symmetry-breaking instanton effects. In Section~\ref{sec:breakingSU9} we will discuss the breaking of quark color-flavor unification, and in Section~\ref{sec:breakingSU3} we will discuss the breaking of lepton flavor and the preservation of a discrete gauge symmetry. At each of these stages we will discuss both the perturbative phenomenology and nonperturbative aspects of the physics including instantons, topological defects, and emergent generalized symmetries. In Section~\ref{sec:matching} we will match the ultraviolet and infrared gauge couplings, and finally in Section~\ref{sec:SMsyms} we will discuss the higher-form symmetries of the Standard Model on which we land in the infrared. We conclude in Section~\ref{sec:conclusion}.

\section{Non-invertible Naturalness} \label{sec:NonInvSym}

The quark-lepton color-flavor unified theory is the setting for our work. In Section \ref{sec:uvSU12} we will introduce this theory and thereafter transition down in energy scales while explaining various symmetry-breaking steps. However, the construction of this theory, and the bottom-up understanding of its core mechanism, has been enabled by our understanding of non-invertible symmetry breaking in flavored $Z'$ theories, and so we will first here explain the relevant concepts and applications. Readers eager to see the top-down model can skip ahead to Section \ref{sec:uvSU12} on a first pass, returning to this section when needed for additional perspective on various symmetry breaking effects.

The basic model-building benefit of analyzing non-invertible chiral symmetries is to systematically learn from the infrared about possible ultraviolet instanton effects that may generate certain couplings in the infrared theory. This guidance goes past 't Hooft naturalness \cite{tHooft:1979rat} in not only providing assurance that the small size of a given coupling may be preserved upon RG flow, but also in pointing to a particular ultraviolet mechanism for generating that small coupling in the first place. This strategy has thus been termed `non-invertible naturalness'. Having this signposted direction for instanton effects is especially useful, as they provide one of few known ways to generate exponential, parametric hierarchies between parameters or scales.

In particular these types of non-invertible symmetries can occur when our gauge group admits more instantons on a general manifold than on $\mathbb{R}^4$ (really its one point compactification $S^4$, as is familiar from instanton studies). The crucial aspect of general manifolds is their ability to support magnetic fluxes on compact cycles, and in fact we can skip considering exotic manifolds entirely if we instead study our theory on $\mathbb{R}^4$ with 't Hooft lines inserted \cite{Fan:2021ntg,Csaki:2024ajo,GarciaGarcia:2025uub,GarciaGarcia:2025xqj}. For simplicity we find the former criterion easier to state.

That is, say we are given an infrared $H$ gauge theory with field strength $H^{\mu\nu}$. As normal, to understand the dynamical effects of instantons in breaking chiral symmetries in the infrared $H$ theory, we study the possible values of
\begin{equation}
    I_{\rm IR}(S^4) = \int_{S^4} \frac{\text{Tr}(H \tilde{H})}{16\pi^2},
\end{equation}
the spectrum of instantons that appear around empty flat space. These instantons break chiral symmetries by generating 't Hooft vertices in our theory on $\mathbb{R}^4$.  However, as theorists, we are free to study our theory on whatever sort of background we wish, and in particular to consider the instantons which appear on a more general manifold
\begin{equation}
    I_{\rm IR}(\mathcal{M}^4) = \int_{\mathcal{M}^4} \frac{\text{Tr}(H \tilde{H})}{16\pi^2}.
\end{equation}
To be perfectly clear, we are studying these configurations \textit{not} because we inherently care about what our theory does on $\mathcal{M}_4$, but because such auxiliary calculations can give us as effective field theorists more information about our $H$ gauge theory that can be useful to understand the theory on $\mathbb{R}^4$. In particular, one may often find that as sets, 
\begin{equation}
    I_{\rm IR}(\mathcal{M}^4) \supset I_{\rm IR}(S^4),
\end{equation}
and there are more instantons allowed on a general manifold than dynamically break chiral symmetries in our IR theory on $\mathbb{R}^4$.
The most familiar example is $U(1)$ gauge theory, where $I_{\rm IR}(S^4) = 0$ and there are no Abelian instantons in flat space, but $I_{\rm IR}(\mathcal{M}^4) = \mathbb{Z}$ meaning in general there are integer-valued instantons on a spacetime with a 2-cycle (for example, the torus $T^4$).\footnote{For $U(1)$ since there is no trace over generators we actually mean $I_{\rm IR} = \int F \tilde F / (32 \pi^2)$.} Another pertinent example is the quotient group $SU(N)/\mathbb{Z}_N$, where $I_{\rm IR}(S^4) = \mathbb{Z}$ being the familiar integer instantons of $SU(N)$, but $I_{\rm IR}(\mathcal{M}^4) = \mathbb{Z}/N$ and in general `fractional' instantons are allowed due to the identification of the $\mathbb{Z}_N$ center of $SU(N)$ with the identity. See e.g. \cite{Koren:2024xof} for some review on the basics of thinking about gauge theories with quotient groups, and e.g. \cite{Anber:2021iip,Cordova:2024ypu,Hong:2025qbw,Reece:2023iqn} for insight into fractional instantons.

The technology of generalized symmetries tells us that in these scenarios, the instantons which live only on exotic manifolds have the effect on the theory in $\mathbb{R}^4$ of converting some chiral symmetries into `non-invertible' symmetries (rather than simply breaking them, as the $I_{\rm IR}(S^4)$ instantons do) \cite{Cordova:2022ieu,Choi:2022jqy}. Furthermore, the technical construction of non-invertible chiral symmetries teaches us something special about their explicit breaking. A non-invertible chiral symmetry relies on the IR theory having a magnetic one-form global symmetry, and admitting magnetic monopole defects. 
As the magnetic one-form symmetry of the $H$ gauge theory is essential, this implies that if we go to a UV $G$ gauge theory which breaks the magnetic one-form symmetries of the $H$ gauge theory (for example, by $\pi_2(G/H) \neq 1$ supplying the correct magnetic monopoles of the $H$ theory as dynamical objects, not simply defects), then the non-invertible chiral symmetries of the $H$ theory \textit{must} be broken in the $G$ theory. 

Systematic consideration of these symmetries leads us to the following model-building strategy called `non-invertible naturalness'. When we go to the UV $G$ theory that breaks the magnetic one-form symmetry we will find
\begin{equation}
    I_{\rm UV}(S^4) = I_{\rm IR}(\mathcal{M}^4),
\end{equation}
and we have upgraded the instantons which did not live on empty flat space in the $H$ gauge theory into instantons which do in the $G$ gauge theory! Then in the UV theory these now have the familiar dynamical effects of chiral symmetry-breaking. While discovering in your infrared theory that a coupling is technically natural teaches you that \textit{if} that coupling were small in the ultraviolet it would stay small into the infrared, discovering that a coupling is non-invertibly natural guides you to ultraviolet theories that \textit{explain} why that coupling is small to begin with.

At a basic level the lesson---that you should look for an ultraviolet theory which supplies the `missing' instantons---could have been guessed long ago. However it seems that historically it was not, as we have been able to use this strategy to discover useful instanton effects in minimal extensions of the SM. Furthermore, quite generally the point of understanding the symmetry statement is to learn the systematics, which teaches us concrete strategies for using this symmetry information. For technical details we refer to earlier papers e.g. \cite{Cordova:2022ieu,Cordova:2022fhg,Cordova:2024ypu}. More generally than just this concept, in recent years many works have appeared using various concepts from generalized global symmetries to learn new particle physics e.g. \cite{Brennan:2023kpw,Brennan:2023tae,Cordova:2022qtz,Craig:2024dnl,Sehayek:2026pvu,Aloni:2024jpb,Cheung:2024ypq,vanBeest:2023dbu,vanBeest:2023mbs,Davighi:2019rcd,Davighi:2024zip,Davighi:2024zjp,Davighi:2025awm,Chen:2024tsx,Burgess:2025geh,Suzuki:2026xvf,Brennan:2024sth,Kaloper:2022yiw,Garcia-Valdecasas:2024cqn}.

\subsection{Non-invertible symmetry breaking in the quark sector} 

Models of $Z'$s, in which an additional $U(1)$ symmetry is gauged, offer a natural bottom-up strategy to search for profitable non-invertible symmetries leading to ultraviolet completions in which small couplings are generated by instantons. The most minimal $Z'$ models are those which do not require additional matter, being simply a gauging of a non-anomalous (approximate) symmetry of the SM itself, a subgroup of the continuous global flavor symmetry $SU(3)^5 \times U(1)_{B-L}$ (where of the other 4 $U(1)$s, one has become hypercharge and three have ABJ anomalies under the three factors of the SM, though some discrete factors are preserved \cite{Koren:2022bam}).

In the lepton sector, the guidance from these non-invertible symmetries was used to find ultraviolet completions of lepton flavor $Z'$ models in which either Dirac or Majorana neutrino masses are generated by lepton flavor instantons, and so automatically exponentially suppressed \cite{Cordova:2022fhg}. In particular, in the Dirac case, the simplest model to implement noninvertible symmetry breaking is an $SU(3)_\ell$ gauge theory of lepton flavor unification. In the quark sector, the guidance from these non-invertible symmetries pointed us to an overlooked theory of $SU(9)$ quark color-flavor unification in which the simplest solutions to the strong CP problem are revived in a flavor-symmetric way (the massless quark solution \cite{Cordova:2024ypu}, the visible PQWW axion \cite{Delgado:2024pcv}, and the invisible DFSZ axion \cite{Choi:2026oqz}, all of which face problems in their SM versions). Here we will unify the quark and lepton flavor effects.

In the quark sector this work will implement the massless quark solution to the strong CP problem via non-invertible symmetry-breaking, and we refer to our earlier \cite{Cordova:2024ypu} for a pedagogical introduction to both the problem and this style of solution. 
But we note an important difference in the non-invertible symmetry-breaking strategy from our previous uses. In prior work, proceeding from the bottom up, we have considered the theory of the SM extended by gauging
\begin{equation}
    \left(SU(3)_C \times U(1)_{B_i + B_j - 2 B_k}\right)/\mathbb{Z}_3,
\end{equation}
where $B_i$ refers to baryon number of the $i$th generation and, interestingly, the crucial quotient is only possible because $N_c = N_g$ in the SM. In \cite{Cordova:2024ypu} we first discovered the theory with the quotient can have a non-invertible, flavor-universal $\mathbb{Z}_3 \subset U(1)$ Peccei-Quinn symmetry which acts as, for example, 
\begin{equation}
    Q_i \rightarrow \omega Q_i, \quad \bar u_i \rightarrow \omega^{-1} \bar u_i, \quad \bar d_i \rightarrow \omega \bar d_i,
\end{equation}
where $\omega = e^{2\pi i/3}$ is the $\mathbb{Z}_3$ generator. This non-invertible symmetry exists firstly because the integer instantons from $I_{\rm IR}(S^4)$ do not fully break the chiral $U(1)$ symmetries as their anomaly coefficients are a multiple of $3 = N_g$. And secondly due to `fractional instantons' present in the theory from $I_{\rm IR}(\mathcal{M}^4)$ which wind through both factors as a result of the non-trivial quotient, and as a result have anomaly coefficients without the factor of $N_g$. Then this $\mathbb{Z}_3$ factor is not broken, but it is converted by the fractional instantons into a non-invertible symmetry.  
The physical effect of this symmetry, in the case with only the SM matter fields, is that it protects all of the down quark yukawas, and turns them into spurions for a non-invertible symmetry. 

We have used this infrared guidance for model building through flavor unification to color-flavor unification
\begin{equation} 
\frac{SU(3) \times U(1)}{\mathbb{Z}_3} \subset \frac{SU(3)\times SU(3)}{\mathbb{Z}_3} \subset SU(9),
\end{equation}
where the non-invertible $\mathbb{Z}_3$ chiral symmetry persists in the intermediate stage. It is not until the color-flavor unification that the color-flavor magnetic monopoles are introduced from $\pi_2\left(SU(9)/\left[SU(3)^2/\mathbb{Z}_3\right]\right) = \mathbb{Z}_3$ and non-invertible symmetry breaking takes place. From the ultraviolet perspective this non-invertible symmetry breaking occurs via the instantons of the $SU(9)$ theory. Then in this UV theory we may start with solely the up-type yukawa and a good PQ symmetry, and let the $SU(9)$ instantons generate the down-type yukawas, bearing out the massless quark solution to strong CP.

In the present case, we can make use of the theoretical data of this non-invertible symmetry in a different way to effectively supply the missing instantons that lower the anomaly coefficient by a factor of three: deconstruction of color by flavor. 
Here we consider a different intermediate completion of the quark flavored $Z'$. The infrared theory can arise from 
\begin{equation}\small
    \frac{SU(3)_1 \times SU(3)_2 \times SU(3)_3 \times U(1)_{B_1 - B_2} \times U(1)_{B_1 + B_2 - 2 B_3} }{\mathbb{Z}_3 \times \mathbb{Z}_3},
\end{equation}
where the quarks of the $i$th generation are charged under $SU(3)_i$. Deconstruction also forbids mixing between generations at this stage, as did turning on the non-Abelian flavor symmetry above, which must be generated upon Higgsing back to the flavor-universal $SU(3)_C$. 

The theory with non-trivial `global structure' can be seen to result from taking the $SU(3)^3 \times U(1)^2$ theory and quotienting the gauge group by a pair of $\mathbb{Z}_3$ elements, each of which lies in the diagonal between one of the Abelian groups and the non-Abelian groups under which those charged quarks transform. This imposes two equivalence relations between elements of $SU(3)^3 \times U(1)^2$,
\begin{align}
    \mathbb{Z}_3:& \mathds{1} \sim (\omega \mathds{1}_1, \omega^{-1} \mathds{1}_2, \mathds{1}_3, \omega^{-1} \mathds{1}, \mathds{1}) \\
    \mathbb{Z}'_3:& \mathds{1} \sim (\omega \mathds{1}_1, \omega \mathds{1}_2, \omega \mathds{1}_3, \mathds{1},  \omega^{-1} \mathds{1}).    
\end{align}
These quotients modify the one-form global symmetries of the theory, as will be discussed further in Section \ref{sec:breakingSU9}: The charged matter content is restricted, and the magnetic monopoles of the theory now include both Abelian and non-Abelian fluxes.
This type of structure fits into a family of gauge groups $\left(SU(N)^N \times U(1)^{N-1}\right)/\mathbb{Z}_N^{N-1}$, which in general can be UV completed into $SU(N^2)$. Again, the possibility of the SM bearing this out is a special consequence of the fact that $N_g = N_c$. 

In this case, by going to the deconstruction phase, the non-invertible symmetry will be broken, but not due to the inclusion of the relevant magnetic monopoles. As discussed above, the non-invertible symmetry here relied on the fact that $SU(3)_C$ is flavor-universal, such that mixed anomalies with it are proportional to $N_g$. Deconstruction allows a novel way to find non-invertible symmetry breaking by explicitly separating the degrees of freedom into $N_g$ different gauge groups. Then the different $SU(3)_i$ integer instantons directly have smaller anomaly coefficients by a factor $1/N_g$, and so accomplish more breaking of the global symmetries than did the flavor-universal $SU(3)_C$. In this theory we still find $I_{\rm intermediate}(\mathcal{M}^4) \supset I_{\rm intermediate}(S^4)$, and the fractional instantons are still missing. However, now the integer instantons of $SU(3)_i$ break $U(1)_{\rm PQ}$ entirely, with no part of it remaining as a non-invertible symmetry. This novel strategy to break non-invertible symmetries will be discussed further in \cite{Hong:2026a}.

Very interestingly, despite new monopoles not appearing at this ultraviolet scale, there is a sense in which this mechanism for breaking non-invertible symmetries by deconstruction still does add in  fractional instantons. This is due to the non-trivial matching of the gauge couplings along this breaking
\begin{equation}
    \frac{1}{\alpha_s} = \sum_{i=1}^3 \frac{1}{\alpha_i}.
\end{equation}
Thus in the flavor-symmetric limit the instanton action \textit{does} jump by a factor of three
\begin{equation}
    S_{\rm inst}^{(\rm intermediate)} =  \frac{1}{3} S_{\rm inst}^{(\rm IR)}.
\end{equation}
We find it intriguing that despite non-invertible symmetry breaking having not technically been accomplished by the existence of true fractional instantons (with fractional topological charge), in practice it is nonetheless accomplished by instantons with fractional action, just like one would find from instantons with fractional topological charge. It would be very interesting to further understand the relationship between these approaches.

\section{The Ultraviolet $SU(12)\times SU(2)_L \times U(1)_R$ Theory} \label{sec:uvSU12}

{\setlength{\tabcolsep}{0.6 em}
\renewcommand{\arraystretch}{1.3}
\begin{table}[t]\centering
\normalsize
\begin{tabular}{|c|c|c|c||c|c|}  \hline
 & $ SU(12)  $ & $ SU(2)_L $ & $ U(1)_R $ & $U(1)_{\rm PQ}$ & $U(1)_{Q+L}$\\ \hline
$\Psi$ & $12$ & $2$ & $0$ & $0$ & $+1$ \\ \hline
$\psi_u$ & $\overline{12}$ & $-$ & $+1$ & $0$ & $-1$ \\ \hline
$\psi_d$ & $\overline{12}$ & $-$ & $-1$ & $+1$ & $-1$ \\ \hline
\end{tabular}\caption{Gauge and global symmetry charges of fermions in the UV phase.} \label{tab:fermionsUV}
\end{table}}

We begin with $SU(3)_\ell$ lepton flavor unified with $SU(9)$ quark color-flavor into $SU(12)$. Gauged $U(1)_{Q-N_cL}$ appears automatically as a subgroup, and relatedly the gauged $U(1)$ symmetry starts as $U(1)_R$ as is familiar from Pati-Salam models. Here we choose not to gauge the full $SU(2)_R$ because a key ingredient will be a PQ symmetry which distinguishes up and down. 

This means the SM fermions are in three irreducible representations (see Table~\ref{tab:fermionsUV}), and we introduce two quark-lepton color-flavor scalar irreps as well in addition to the Higgs (see Table~\ref{tab:scalarsUV}): 
\begin{itemize}
    \item The \textit{complex} adjoint $\Sigma^a_b$ which spontaneously breaks the $SU(12)$. We outline this in stages in Figure~\ref{fig:BreakingChart} for conceptual benefit but hierarchical breaking is not necessary, and indeed our simple benchmark will join most of the steps. This is simply an organization of two real adjoints and the $U(1)_\Sigma$ need not be respected---indeed, to generate the weak CP-violating angles $\Sigma$ should spontaneously and/or explicitly violate this symmetry.
    \item The two-index symmetric $\phi^{ab}$ which is the only new scalar with an electroweak quantum number, and can interact with the SM fermions at tree-level. Remarkably, this plays multiple distinct roles: spontaneously breaking lepton flavor and providing right-handed neutrino Majorana masses, spontaneously breaking $U(1)_R$ and $U(1)_{Q-N_cL}$ down to hypercharge, preserving a discrete lepton flavored gauge symmetry which prevents proton decay, and also dynamically distinguishing up-type vs down-type quarks to enable the generation of the CKM phase.
\end{itemize}

We start with an extremely simple Lagrangian
\begin{equation} \label{eqn:SU12lag}
    \mathcal{L} = y_t H \Psi \psi_u + \half \lambda \phi \psi_u \psi_u + V(H,\Sigma,\phi),
\end{equation}
with $y_t$ and $\lambda$ two free parameters which are naturally $\mathcal{O}(1)$. We do not fully specify the potential through which the scalars interact, but take it as given that the scalars can condense in the directions we dictate, as is usual in initial work on theories of unification.

The theory of Eqn.~\ref{eqn:SU12lag} additionally has two global symmetries which act nontrivially on the fermions, both of which are anomalous. The absence of interactions for $\psi_d$ is enforced by a classical symmetry effecting $\psi_d \rightarrow e^{i\alpha} \psi_d$, which at the quantum level becomes a Peccei-Quinn symmetry and will be crucial for the instanton effects we will find. As to the four fields which do interact in Eqn.~\ref{eqn:SU12lag}, the free theory of these fields would have $U(1)^4$, two directions of which are broken by the two interactions, leaving both the gauged $U(1)_R$ and a global $U(1)_{Q+L}$ in addition to the $U(1)_{\rm PQ}$. In Table~\ref{tab:fermionsUV} we take a useful choice of basis for the global symmetries in which we identify $U(1)_{\rm PQ}$ as a direction with solely $SU(12)$ mixed anomaly and $U(1)_{Q+L}$ as a direction with solely $SU(2)_L$ mixed anomaly. We refer to our recent \cite{Choi:2026oqz} for detailed discussion on the abstract space of global symmetries and the scalar field space in theories with PQ symmetries.  

{\setlength{\tabcolsep}{0.6 em}
\renewcommand{\arraystretch}{1.3}
\begin{table}[t]\centering
\normalsize
\begin{tabular}{|c|c|c|c||c|c| }  \hline
 & $ SU(12)  $ & $ SU(2)_L $ & $ U(1)_R $ & $U(1)_{\rm PQ}$ & $U(1)_{Q+L}$ \\ \hline
$H$ & $-$ & $2$ & $-1$ & $0$ & $0$ \\ \hline
$\Sigma^a_b$ & $143$ & $-$ & $0$ & $0$ & $0$ \\ \hline
$\phi^{ab}$ & $78$ & $-$ & $-2$ & $0$ & $+2$\\ \hline
\end{tabular}\caption{Gauge and global symmetry charged of scalars in the UV phase.}\label{tab:scalarsUV}
\end{table}}

The direction $U(1)_{Q+L}$ is quark plus lepton number, which parallels the phenomenology of baryon plus lepton number in the Standard Model. This symmetry is `accidental' in that any renormalizable Lagrangian must classically respect it due to the gauge charges. The $SU(2)_L$ instantons already give the familiar SM 't Hooft vertex between 12 fermions $\mathcal{L} \subset \epsilon^{a_1 \dots a_{12}} \Psi_{a_1} \dots \Psi_{a_2}$ which allow processes that exchange 9 quarks for 3 leptons, and breaks $U(1)_{Q+L}$ into the conserved $\mathbb{Z}_{12}^{Q+L}$.

The $SU(12)$ instantons fully violate $U(1)_{\rm PQ}$, and as these effects grow in magnitude into the infrared they will lead to the 't Hooft vertices and noninvertible symmetry breaking we need to generate the missing yukawas and solve strong CP. This holds true because the $SU(12)$ gauge coupling is asymptotically free at $\beta_{12} = -109/3$. On the other hand the $U(1)_R$ factor has $\beta_R = 362/3$ as the result of the large scalar $78$ representation, 
implying that a UV completion must be relatively close. Here this will not impose any requirements---as we will discuss further in the next section, the scale of our dynamics will necessarily be quite high already, so that $M_{\rm pl}$ will cut off the $U(1)_R$ theory before the Landau pole (see Figure \ref{fig:RunningCouplings}).

The matter content of this theory is very minimal, and we will end up finding that some colored spectator fields are needed to modify the running and consistently achieve the many things we are demanding from this minimal setup. All that matters about these spectator fields is their contribution to the beta function, so we are agnostic about their identities and here assume they have no interactions with the SM fields. However, we mention one well-motivated way to add colored degrees of freedom is to explore upgrading the Higgs sector from a color-flavor singlet to include $SU(12)$ singlet + adjoint $H_u$ and $H_d$, which can achieve the modified running, can help facilitate full UV gauge coupling unification, and can also help generate the full yukawa structure. We postpone analyzing this structure to future work, and in Section~\ref{sec:matching} just take as a free parameter an additional contribution $\beta_{\rm spectator}$ coming from the spectator fields.

\subsection{Global structure and generalized symmetries}

The ultraviolet theory contains a continuous Abelian factor, so this theory possesses a $U(1)^{(1)}_m$ magnetic one-form symmetry under which the 't Hooft lines are charged. This is due to the absence of dynamical magnetic monopoles for $U(1)_R$, and this magnetic one-form symmetry will match on to the $U(1)_Y$ hypercharge magnetic one-form symmetry\footnote{In the Standard Model phase this hypermagnetic one-form symmetry is intertwined in an approximate two-group with the SM flavor symmetries \cite{Cordova:2022qtz}. In that paper the connection with traditional, `vertical' GUTs was explored; it would be very interesting to understand how that structure emerges from a flavorful GUT like this one.} and then the $U(1)_{\rm EM}$ electromagnetic one-form symmetry. This means that this theory itself does not suffer a monopole problem, as electromagnetic monopoles are not present. We will see below that other magnetic monopoles will emerge at intermediate stages, but these will all get confined by strings before the SM phase is reached. In a given further ultraviolet theory---for example embedding $U(1)_R \subset SU(2)_R$, or embedding the whole theory in a five-dimensional orbifold---there will be extra topology which will also provide dynamical electromagnetic monopoles and break the one-form symmetry explicitly.  
But this is not the only consideration for the one-form symmetries of this UV theory.

There are two possible versions of this theory which possess either a $\mathbb{Z}_2^{(1)}$ electric one-form symmetry, or a modified $U(1)^{(1)}$ magnetic one-form symmetry. Consider a simultaneous transformation by $\mathbb{Z}_2$ in each of the three gauge factors---this diagonal $\mathbb{Z}_2$ acts trivially on all the fields listed in Tables~\ref{tab:fermionsUV} and~\ref{tab:scalarsUV}. Then we are allowed to consider two possibilities for the gauge group.

Option 1: $G_{\rm UV} = SU(12) \times SU(2)_L \times U(1)_R$.  \ This theory has a $\mathbb{Z}_2^{(1)}$ electric one-form symmetry under which the charged Wilson lines are those that cannot be ended on the matter included in this theory so far---that is, any Wilson line which picks up a negative sign under that shared $\mathbb{Z}_2$ transformation. In the far infrared, this maps on to Wilson lines with fractional electric charge $(1/2 + k)e$, with $k \in \mathbb{Z}$. The minimal field content above does not include any such fields, being only a theory of the chiral SM matter. But for this gauge group, with general matter there will be additional complex scalars or Dirac fermions on which those Wilson lines can end because they do transform with a negative sign under that $\mathbb{Z}_2$. The masses of these fields can be entirely separate from the scales generated for the SM fields in this theory, and in particular any mass for such a fermion is technically natural. If they happen to be light, these FCPs may be searched for at the LHC \cite{Koren:2024xof,Koren:2025utp}. Quantum gravity suggests higher-form global symmetries should not be absolute, so if this is the gauge group one should generically expect such additional fields to be present. This gauge theory will flow in the infrared to the SM electroweak gauge group with no global structure $G_{\rm EW} = SU(2)_L \times U(1)_Y$.

Option 2: $G'_{\rm UV} = (SU(12) \times SU(2)_L \times U(1)_R)/\mathbb{Z}_2$. This theory may be constructed by gauging the $\mathbb{Z}_2^{(1)}$ electric one-form symmetry of the theory above. As the fractionally-charged Wilson lines are charged under this one-form symmetry, gauging it removes them from the spectrum of gauge-invariant line operators and as a result FCPs cannot exist in this theory. Group-theoretically the magnetic one-form symmetry remains $U(1)^{(1)}_m$ as $\pi_1(G'_{\rm UV}) \simeq \pi_1(G_{\rm UV})$, but now the closed paths through the gauge group are changed. The fact that the minimal winding path now goes from the origin to $(-\mathds{1}_{12}, -\mathds{1}_2, e^{i\pi})$ means that the $R$-magnetic monopoles with odd charges are now dressed also with half-fractional non-Abelian magnetic fluxes. In the infrared this theory will flow to the electroweak group $G_{\rm EW} = (SU(2)_L \times U(1)_Y)/\mathbb{Z}_2$.

The physics that we discuss in the rest of this work will not be affected by the choice of $\mathbb{Z}_2^{(1)}$ symmetry. However, connecting this theory to a further ultraviolet completion may demand one or the other choice. 

We comment briefly on the existence of non-invertible symmetries in this phase: There aren't any, despite that there are `missing' instantons $I(S^4) \subset I(\mathcal{M}^4)$. On a general manifold, there are integer instanton backgrounds of $U(1)_R$ which break the $U(1)_{\rm PQ}$ Peccei-Quinn symmetry to $\mathbb{Z}^{\rm PQ}_{12}$, but as this symmetry is already fully anomalous in $SU(12)$ integer instanton backgrounds, this does not imply a non-invertible symmetry unless we were to turn off the $SU(12)$ gauge coupling.  A more subtle possibility to check is that in the theory with the $\mathbb{Z}_2$ quotient there are \emph{fractional} instantons which have correlated, fractional instanton numbers under all three groups. However, these do not violate $U(1)_{Q+L}$ (which can be understood due to their inherent left-right symmetry inherited from the possible $SU(2)_R$ UV completion) so also do not lead to non-invertible symmetries in this theory. In a theory that introduced more fields and so more global symmetries, it is possible that these Abelian integer instantons or fractional instantons would lead to interesting non-invertible symmetries, pointing to nonperturbative physics effects in a further full unification into a simple group. But here we make no use of these instantons.

\section{Breaking $SU(12)$ Separates Quark and Lepton Flavor} \label{sec:breakingSU12}

One structural mystery of the Standard Model is that the global quark and lepton flavor symmetries are broken in very different ways, from the relative sizes of the masses to the mixing patterns. If we are to upgrade flavor to a gauge symmetry in the context of a theory that unifies quarks and leptons, this seems to be an obstruction.

While it is possible to break $SU(12) \rightarrow SU(4)_C \times SU(3)_F$ leading to a Pati-Salam(-esque) theory with gauged flavor,  it turns out to be more profitable to break `in the Pati-Salam direction' first 
\begin{equation}
    SU(12)\rightarrow \frac{SU(9)_{\rm quarks} \times U(1)_{Q - N_c L}}{\mathbb{Z}_3} \times SU(3)_{\rm  \ell eptons}.
\end{equation} This leads to a theory of \textit{separate} gauged flavor symmetries for the quarks and the leptons, which is crucial to the interesting phenomenologies we will find. The non-trivial global structure can be seen from the fact that the generator of $Q-N_cL$ is the familiar $T^{15}$ of $SU(4)$ in Pati-Salam, but now extended across $SU(12)$ as the direct product $\mathds{1}_3 \otimes T^{15}$, with $\mathds{1}_3$ the $3\times 3$ identity matrix. 
After breaking this theory down to the SM, this quotient will eventually become the $\mathbb{Z}_3$ quotient of the SM, $(SU(3)_C \times U(1)_Y)/\mathbb{Z}_3$, as is familiar from Pati-Salam theories.

This breaking can be implemented by the adjoint $\Sigma^a_{\ b}$ condensing with vev 
\begin{equation}
    \langle\Sigma^a_{\ b} \rangle = \Lambda_{12}\text{diag}(3 \mathds{1}_3,-\mathds{1}_9).
\end{equation}
One of the off-diagonal `bifundamentals' $(9, \overline{3})_{+4}$ is eaten by the gauge bosons and becomes heavy. We show in Tables~\ref{tab:fermionsStep1} and~\ref{tab:scalarsStep1} the charges of the matter under the remaining gauge symmetries.

{\setlength{\tabcolsep}{0.6 em}
\renewcommand{\arraystretch}{1.3}
\begin{table}[t]\centering
\small
\begin{tabular}{|c|c|c|c|c|c|}  \hline
 & $ SU(9)_{q}  $ & $ SU(3)_{\ell}  $ & $ SU(2)_L $ & $ U(1)_R $ & $U(1)_{Q-N_c L}$ \\ \hline
$\Psi_q$ & $9$ & $-$ & $2$ & $0$ & $+1$ \\ \hline
$\Psi_\ell$ & $-$ & $3$ & $2$ & $0$ & $-3$ \\ \hline
$\psi_{qu}$ & $\overline{9}$ & $-$ & $-$ & $+1$ & $-1$ \\ \hline
$\psi_{qd}$ & $\overline{9}$ & $-$ & $-$ & $-1$ & $-1$  \\ \hline
$\psi_{\ell u}$ & $-$ & $\overline{3}$ & $-$ & $+1$ & $+3$ \\ \hline
$\psi_{\ell d}$ & $-$ & $\overline{3}$ & $-$ & $-1$ & $+3$ \\ \hline
\end{tabular}\caption{Fermions in the $SU(9) \times SU(3)$ phase and their gauge representations.}\label{tab:fermionsStep1}
\end{table}}

{\setlength{\tabcolsep}{0.6 em} 
\renewcommand{\arraystretch}{1.3}
\begin{table}[h!]\centering
\small
\begin{tabular}{|c|c|c|c|c|c|}  \hline
 & $ SU(9)_{\rm q}  $ & $ SU(3)_{\rm \ell}  $ & $ SU(2)_L $ & $ U(1)_R $ & $U(1)_{Q-N_c L}$ \\ \hline
$H$ & $-$ & $-$ & $2$ & $+1$ & $0$ \\ \hline
$\Sigma^a_b$ & $80$  & $-$ & $-$ & $0$ & $0$ \\ \hline
$\Sigma^\alpha_\beta$ & $-$ & $8$ & $-$ & $0$ & $0$ \\ \hline
$\cellcolor{lightgray} \Sigma^a_\beta$ & $9$ & $\bar 3$ & $-$ & $0$ & $+4$ \\ \hline
$\cellcolor{lightgray} \Sigma$ & $-$ & $-$ & $-$ & $0$ & $0$ \\ \hline
$\cellcolor{lightgray} \phi^{ab}$ & $45$ & $-$ & $-$ & $-2$ & $+2$ \\ \hline
$\phi^{a\alpha}$ & $9$ & $3$ & $-$ & $-2$ & $-2$ \\ \hline
$\phi^{\alpha\beta}$ & $-$ & $6$ & $-$ & $-2$ & $-6$ \\ \hline

\end{tabular}\caption{Complex scalars in the $SU(9) \times SU(3)$ phase. The fields in gray cells are not needed at lower energies and could receive large $SU(12)$-breaking sized masses.  $a,b = 1..9$ are $SU(9)$ indices and $\alpha, \beta = 1..3$ are $SU(3)$ indices.}\label{tab:scalarsStep1}
\end{table}}

The Lagrangian decomposes to
\begin{align}
    \mathcal{L} &= y_t H \left(\Psi_q \psi_{qu} + \Psi_\ell \psi_{\ell u} \right) \\ &+ \lambda \left( \phi^{ab} \psi_{qu,a} \psi_{qu,b} + \phi^{a\alpha} \psi_{qu,a} \psi_{\ell u,\alpha} + \phi^{\alpha\beta} \psi_{\ell u,\alpha} \psi_{\ell u,\beta}\right), \nonumber
\end{align}
and higher-dimension operators are generated upon integrating out the heavy gauge bosons such as
\begin{equation} \label{eqn:higherDimPQ}
    \Lambda_{12}^2 \mathcal{L}_6 \simeq (\psi_{q d} \sigma^\mu \psi_{\ell d}^\dagger) (\psi_{\ell d} \sigma_\mu \psi_{q d}^\dagger) + \dots
\end{equation}
arising from the currents coupling to the heavy $(9,\bar 3)_{+4}$ vector boson.

Naively the quarks and leptons may enjoy distinct vector-like global symmetries for quark and lepton number, but these are linked by the gauged $U(1)_{Q-N_cL}$. We may continue to consider quark plus lepton number which the $SU(2)_L$ anomaly continues to break to $\mathbb{Z}_{12}^{Q+L}$. We could alternatively consider, for example, $Q + N_c L = 3(B+L)$ which is broken by the anomaly instead to $\mathbb{Z}_{18}^{Q+N_cL}$ as happens in the low-energy Standard Model as well and protects the proton \cite{Koren:2022bam}. There are also PQ symmetries acting separately on the right-handed down-quarks or charged leptons, each of which is fully broken by $SU(9)$ or $SU(3)_\ell$ mixed anomalies. The irrelevant operator in $\mathcal{L}_6$ couples them together and breaks them down to their difference $U(1)_{\rm PQ, q - \ell}$, which has the effect of placing a lower bound on the amount of Peccei-Quinn breaking in one sector with respect to the other (see Footnote~\ref{foot:higherDimPQ} below).

In this phase we also have an emergent $U(1)^{(1)}$ magnetic one-form global symmetry due to the Abelian factor of $U(1)_{Q-N_cL}$. The quotient between $Q-N_cL$ and $SU(9)$ again modifies the structure of this one-form symmetry, as the minimal winding path now goes from the origin to $(\omega^{-1} \mathds{1}_9, \omega)$ with $\omega = \exp i 2\pi/3$. Then among the $\mathbb{Z}$-spectrum of 't Hooft lines, those not at $3 \mathbb{Z}$ units of $Q - N_cL$ magnetic flux have additional compensating one-third fractional $SU(9)$-magnetic fluxes as well. If $SU(12)$ is dynamically broken in the early universe, these 't Hooft lines are brought to life as magnetic monopoles which form necessarily due to causality \cite{Kibble:1976sj}. It would be very interesting to understand if these monopoles have a Callan-Rubakov-like effect \cite{Callan:1982au,Rubakov:1982fp} due to the mixed anomaly of $SU(9)$ with the PQ symmetry. Such monopoles could potentially play a role in baryogenesis \cite{Brennan:2024sth} with the additional benefit that at a lower scale when $U(1)_{Q-N_cL}$ is Higgsed, they will become endpoints of cosmic strings whose tension will cause them to annihilate.

\subsection{Beta functions and matching}

The running of the gauge couplings in this stage depends on which degrees of freedom get mass at the scale $\Lambda_{12}$, so there is a range of possible beta functions for each factor. Some of the scalars could receive a large mass $\sim \Lambda_{12}$ and be removed from the spectrum, while others are needed at subsequent breaking steps. The adjoints are needed to spontaneously break the flavor gauge groups, as is the two-index symmetric lepton-flavored $\phi^{\alpha\beta}$, while $\phi^{a\alpha}$ is used in the flavor scheme to distinguish up vs down. The others are not strictly necessary, and our benchmark model decouples them.

Across the $SU(12)$-breaking scale the non-Abelian gauge couplings match as they have trivial index of embedding, $g_{\rm 12} = g_{9} = g_\ell$. However for the Abelian factor the sensible infrared quantization of the couplings often deviates from the ultraviolet normalization of $\text{Tr}(T^2) = \half$ over the adjoint. We have the non-trivial matching condition $g_{\rm 12} = \sqrt{72} g_{Q-N_cL}$.\footnote{Note this large factor is because we are normalizing $Q-N_cL$ in the unbroken phase with the quarks having charge 1, which is thrice their $B-L$ charge. In the standard normalization this would be $g_{\rm 12} = \sqrt{8} g_{B-L}$. Note also that this differs by a factor of root three from the relation in Pati-Salam theories, $g_4 = g_3 = \sqrt{8/3} g_{B-L}$, as the gauge groups are thrice as big.} 
In terms of running, the beta functions for both gauge flavor symmetries remain asymptotically free at $\beta_9 = -169/6$ for $SU(9)$ and $\beta_\ell = -19/3$ for $SU(3)_\ell$, which means as we flow down toward the IR, $g_{9}$ grows with respect to $g_\ell$. This is crucial as it allows the different running of couplings to be responsible for generating the difference between the bottom and the tau yukawa (along possibly with different scales for the breaking of these symmetries) as we will discuss in the next section. 
For the $U(1)$ factors, our benchmark has $\beta_{Q-N_cL} = 204$ and $\beta_R = 182/3$.

\begin{figure}
    \centering
    \includegraphics[width=.7\linewidth]{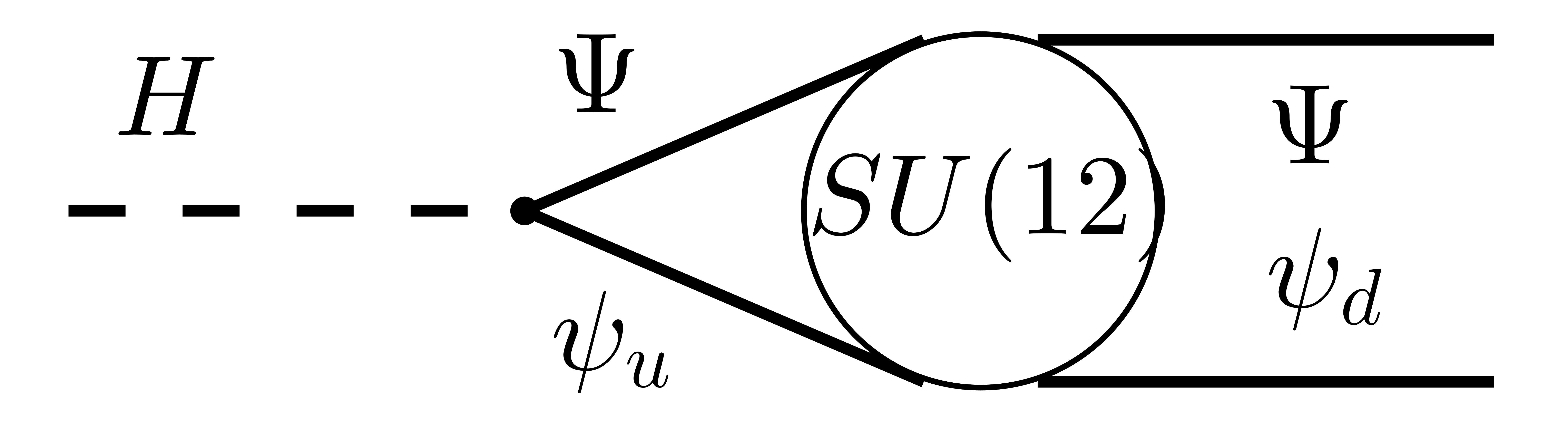}
    \caption{An $SU(12)$ instanton violates PQ by one unit and generates a down-type yukawa from an up-type yukawa. The largest contributions come after $SU(12)$ breaks and there are separate $SU(9)$ instantons for the quarks and $SU(3)_\ell$ instantons for the leptons, but the form remains the same.}
    \label{fig:QLCFSU12Instanton}
\end{figure}

\subsection{Instanton effects and non-invertible symmetry breaking}

From the perspective of the ultraviolet theory, we have simply the breaking of anomalous symmetries by instantons, which for an asymptotically free theory dominantly occurs in the infrared. The infrared non-invertible symmetries are broken in the $SU(9)$ and $SU(3)_\ell$ gauge theories by quark color-flavor and lepton flavor instantons, which generate separately both of $y_b$ and $y_\tau$. In principle we would like the small hierarchy $y_\tau/y_b$ to result from slight differences in the evolution of the two sectors following the breaking of the unified group. See Figure~\ref{fig:QLCFSU12Instanton} showing the generation of the down-type yukawas from the 't Hooft vertices.

The 't Hooft vertices of $SU(3)_\ell$ have four fermi zero-modes, one each of $\psi_{\ell u} = \bar \nu_\alpha$ and $\psi_{\ell d} = \bar e_\alpha$, and two for $\Psi_\ell = L^\alpha$. Using the tree-level instanton action $8 \pi^2/g^2$ evaluated at the scale of the largest breaking, after the up-type yukawa is used to contract the neutrino legs the instantons generate, at the lowest order of approximation
\begin{equation}
    \mathcal{L} \supset y_t H L^\alpha \bar \nu_{\alpha} + y_t^\star e^{- \frac{2\pi}{\alpha_\ell(\Lambda_{\ell})}} \tilde{H} L^{\alpha} \bar e_{\alpha},
\end{equation}
where $\Lambda_\ell$ is the scale at which $SU(3)_\ell$ breaks, and the needed coupling to produce $y_\tau$ is large enough that it is not clear the dilute instanton gas approximation can be trusted, and precise predictions for the needed $\alpha_\ell$ coupling in this theory will require lattice simulation. To be clear, it is only the size of the coefficient which is uncertain---the generation of this operator is a robust prediction due to the ABJ anomaly structure.   

The 't Hooft vertices of $SU(9)$ analogously have four fermi zero-modes as well, one each of $\psi_{qu} = \bar u_{a}$ and $\psi_{qd} = \bar d_{a}$, as well as two for $\Psi_q = Q^{a}$. We again contract the up-type legs with the up-type yukawa, now producing a down-type yukawa of the form

\begin{equation}
    \mathcal{L} \supset y_t H Q^{a} \bar u_{a} + y_t^\star e^{- \frac{2\pi}{\alpha_9(\Lambda_9)}} \tilde{H} Q^{a} \bar d_{a},
\end{equation}
where $\Lambda_9$ is the scale at which $SU(9)$ breaks, and the needed coupling here is even larger, so again lattice simulations will be required to determine the necessary size of $\alpha_9$ to generate the correct yukawa, just as they were needed to understand the original massless quark solution.\footnote{From here we can see the effect of the higher-dimensional operators in Eqn~\ref{eqn:higherDimPQ} is to generate the PQ-violating leptonic operator from the $SU(9)$ 't Hooft vertex, but suppressed by $\Lambda_9^2/\Lambda_{12}^2$. This could be a problem if we wanted to generate only $y_\tau \ll y_b$, but in fact they are not so different and so the higher-dimensional operators are not dangerous. \label{foot:higherDimPQ}}\footnote{An important difference from the original massless quark solution and from our earlier massless quark solution in quark color-flavor unification \cite{Cordova:2024ypu} is that the dynamics here take place at a high scale, as they are tied to the seesaw scale. While the massless quark solution can be automatically safe from the quality problem if all the scales are $ \lesssim 10^8$ GeV, here our model does suffer from the quality problem. This is one reason to search for a quark-lepton flipped model from extra-dimensional dynamics in which one begins with $y_t$ and $y_\tau$ and generates $y_b$ and $y_\nu$, which could take place at much lower scales. Or we must add a module to solve the quality problem, but we leave study of this important problem to future work.}

The quark and lepton theories begin with equal gauge couplings $\alpha_\ell(\Lambda_{12}) = \alpha_{\rm 9}(\Lambda_{12})$, but the different multiplicity factors lead to different one-loop running. If we focus on the ratio $y_\tau/y_b$, then the boundary condition at $\Lambda_{12}$ drops out and this ratio depends on the possible mild amount of scale separation between the quark and lepton flavor-breaking scales. Proceeding with the naive estimate 
\begin{equation}
    y_\tau/y_b = \exp 2\pi \left(\alpha_{9}^{-1}(\Lambda_9)-\alpha_{\ell}^{-1}(\Lambda_{\ell})\right)
\end{equation}
and since
\begin{align}
    \alpha_9^{-1}(\Lambda_9) &\simeq \alpha^{-1}(\Lambda_{12}) - \frac{\beta_{9}}{2\pi} \log\left(\frac{\Lambda_9}{\Lambda_{12}}\right) \\
    \alpha_{\ell}^{-1}(\Lambda_{\ell}) &\simeq \alpha^{-1}(\Lambda_{12}) - \frac{\beta_{\ell}}{2\pi} \log\left(\frac{\Lambda_{\ell}}{\Lambda_{12}}\right)
\end{align}
then we can find the needed lepton-breaking scale as
\begin{equation}
    \frac{\Lambda_{\ell}}{\Lambda_{12}} = \left(\frac{y_\tau}{y_b}\right)^{1/\beta_\ell} \left(\frac{\Lambda_{9}}{\Lambda_{12}}\right)^{\beta_{9}/\beta_\ell}.
\end{equation}

The structure of our model has type-1 seesaw Majorana masses generated at $SU(3)_\ell$ breaking which fixes the overall scale, $\Lambda_{\ell}\sim 10^{15}$ GeV. If we choose the two flavor-breaking scales equal, then all three scales must be nearly degenerate $\Lambda_\ell = \Lambda_{9} \simeq 0.96 \Lambda_{12}$. 
If we want a separation of scales, the most we can do is raising $SU(12)$ breaking to around the Planck scale $\Lambda_{\rm \ell}/\Lambda_{12} \sim 10^{-3}$ which requires $\Lambda_{9}/\Lambda_{12} \sim 1/5$. Any choice works fine at the level we're considering, but for simplicity in Section~\ref{sec:matching} we will consider the case where all the breaking occurs at just around the same scale.

While the $\alpha$-dependence drops out of this estimate and so one may hope this prediction is more robust than the absolute values of $y_\tau, y_b$, there is much physics they ignore. 
These estimates could be made more precise by accounting for the instanton density in further detail (see e.g. \cite{Csaki:2023ziz, Sesma:2024tcd} for recent approaches). As our purpose here is understanding the overall structure, we will not belabor a more-precise instanton computation. We hope this interesting scenario, along with previously studied possibilities for these $SU(9)$ instantons \cite{Cordova:2024ypu,Delgado:2024pcv,Choi:2026oqz}, motivates lattice simulations of this gauge theory.

\section{Breaking Quark Color-Flavor} \label{sec:breakingSU9}

Here we consider further breaking of $SU(9)$ by the adjoint vev
\begin{equation}
    \langle \Sigma^a_{\ b} \rangle = \Lambda_{\rm 9} \text{diag}(+\mathds{1}_3,0 \times \mathds{1}_3,-\mathds{1}_3),
\end{equation} which breaks 
\begin{equation}
    SU(9) \rightarrow \frac{SU(3)_1 \times SU(3)_2 \times SU(3)_3 \times U(1)_{T^3} \times U(1)_{T^8}}{\mathbb{Z}_3 \times \mathbb{Z}_3}
\end{equation}
performing color-flavor deconstruction.\footnote{In prior work we have considered the symmetry-breaking pattern $SU(9) \rightarrow SU(3)^2/\mathbb{Z}_3$ which may be implemented by the three-index symmetric $165$ of $SU(9)$. This pattern is viable here as well, with the most simple possibility having the $165$ arise from the $364$ of $SU(12)$. This results in a $165$ with nontrivial $Q-N_cL$ charge, so it should also carry $U(1)_R$ charge in the right combination such that its condensation preserves $U(1)_Y$. While this is a possibility, the condensation of such a field will break an interesting discrete gauge symmetry and prevent some of the physics we will discuss in the next section, so we will think about another possibility. Producing a $165$ which is neutral under $Q-N_cL$ requires such a large $SU(12)$ representation ($3003$-dimensional) that we do not pursue this possibility.} Under this decomposition we have the branchings
\[
      9 \rightarrow (3,1,1)_{(+1,+1)} \oplus (1,3,1)_{(-1,+1)}  \oplus (1,1,3)_{(0,-2)} \nonumber \]
\begin{align}     80 &\rightarrow 2(1,1,1)_{(0,0)} \nonumber \\ & \qquad \oplus(8,1,1)_{(0,0)}  \oplus (1,8,1)_{(0,0)}  \oplus (1,1,8)_{(0,0)}  \nonumber \\
      &\qquad \oplus  (3,\bar 3,1)_{(+2,0)}  \oplus (1,3,\bar3)_{(-1,+3)}  \oplus (\bar 3, 1, 3)_{(-1,-3)}  \nonumber \\
      &\qquad \oplus (\bar 3,3,1)_{(-2,0)} \oplus (1,\bar 3, 3)_{(+1,-3)} \oplus (3,1,\bar 3)_{(+1,+3)}. \nonumber
\end{align}
Here the subscripts denote the charges under $U(1)_{T^3} \times U(1)_{T^8} = U(1)_{B_1 - B_2} \times U(1)_{B_1 + B_2 - 2 B_3}$, and all representations can be constructed from the fundamental, but we give the adjoint for convenience. The labels come from writing the generators as $\mathbb{I}_3 \otimes T^i$ with $T^i$ a diagonal generator of $SU(3)$.\footnote{Note also that the $SU(3)_i$ labels need not map directly to what we normally think of as the $i$th quark generation. It would also be interesting to understand the possibility of a mismatch in quark and lepton generation numbers arising out of this framework.} 

The structure of the quotients within $SU(9)$ has been discussed already in Section~\ref{sec:NonInvSym}, but we recall that due to further unification our representations also transform nontrivially under $U(1)_{Q-N_cL}$, and the deconstruction phase inherits another $\mathbb{Z}_3$ quotient from the $SU(9)$ quotient. In this phase the quotient is generated by 
\begin{equation}
    \mathbb{Z}_3: \mathds{1} \sim (\omega^{-1} \mathds{1}_1, \omega^{-1} \mathds{1}_2, \omega^{-1} \mathds{1}_3, \omega  \mathds{1})
\end{equation}
as an element of $SU(3)^3 \times U(1)_{Q-N_cL}$.

Furthermore, note that there is also an $S_3$ rearrangement symmetry in this infrared group, which is perhaps more obvious if we choose a different basis of the $U(1)$ groups such as $B_1 - B_2$ and $B_1 - B_3$. It's clear then that both possible family differences are gauged, which is an $S_3$ invariant statement, although in a given basis the $S_3$ mixes up the $U(1)$ symmetries as well. Then the group structure at this stage is actually
\begin{equation}
    \left(\left[\frac{SU(3)^3 \times U(1)^2}{\mathbb{Z}_3 \times \mathbb{Z}_3} \rtimes S_3\right]\times U(1)_{Q-N_cL}\right)/\mathbb{Z}_3 .
\end{equation}
We note that at this intermediate stage there is a further emergent $\left[U(1)^{(1)}\right]^2$ magnetic one-form symmetry in the quark color-flavor sector as a result of the $U(1)^2$ gauged quark flavor difference symmetries. There are 't Hooft lines with magnetic fluxes labeled by $(n_3, n_8) \in \mathbb{Z}^2$ in the $(T^3,T^8)$ directions, but the structure of these is also sensitive to the quotient. In particular, if $n_3, n_8 \in 3 \mathbb{Z}$ then these have purely Abelian magnetic fluxes, but a general monopole also has fractional non-Abelian magnetic fluxes of the three $SU(3)_i$ groups as $(n_3 + n_8, -n_3 + n_8, n_8) \ (\text{mod } 3)$ units of flux. Of course the $SU(9)$ theory explicitly breaks the magnetic one-form symmetries in the UV by providing the dynamical magnetic monopoles on which the 't Hooft lines can end, meaning that in the early universe this breaking leads to further production of magnetic monopoles past those already produced in the $SU(12)$-breaking. These monopoles will also annihilate once the deconstructed theory is broken down further to the strong sector of the SM. 

\begin{figure}
    \centering
    \includegraphics[width=0.5\linewidth]{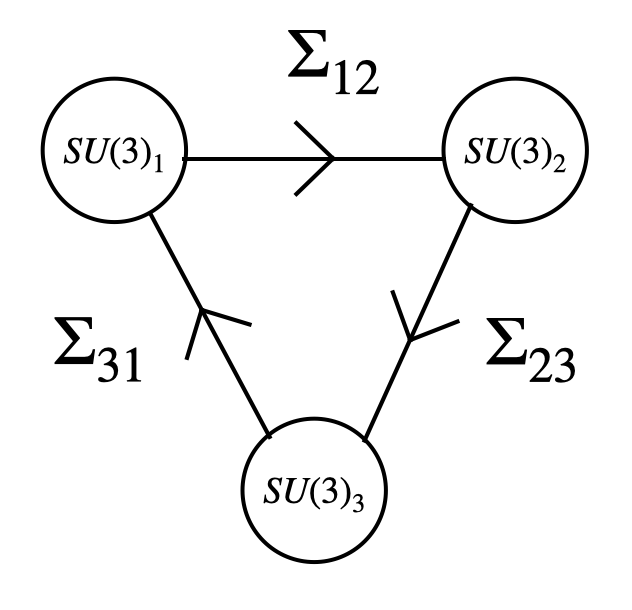}
    \caption{A quiver diagram depicting the non-Abelian parts of the gauge theory, as well as the bifundamental link fields which will break the deconstructed symmetry.}
    \label{fig:Quiver}
\end{figure}

In the breaking from $SU(9)$, half of the bifundamental scalars are eaten by the heavy $SU(9)/SU(3)^3$ vector bosons, but one complex $(3, \bar 3, 1) \oplus (1,3,\bar 3) \oplus (\bar 3, 1, 3)$ remains and we assume these stay light to accomplish the next stage of breaking, see Figure \ref{fig:Quiver}. The condensation of these `link fields' as 
\begin{equation}
    \langle \Sigma^{a_i}_{b_j}\rangle = \Lambda_3 \delta^{a_i}_{b_j}
\end{equation}
breaks the deconstructed gauge theory back to the flavor reunified color group $SU(3)^3 \rightarrow SU(3)_C$. The $U(1)$ factors will be broken simultaneously as a result of their Abelian charges. This step of breaking must happen at a high scale $\gtrsim 1000 \rm \ TeV$ to avoid constraints on FCNCs. We note in recent years there has been interesting work on models of flavor deconstruction where the link fields get vevs at very different scales e.g. \cite{Bordone:2017bld,Davighi:2023iks} and it would be interesting to explore that idea in this context, such that new fields charged under the third-generation factor could be accessible at colliders.

As remarked in the introduction, it would be very interesting to thoroughly understand the physics of the semidirect product rather than the direct one. In breaking to a non-Abelian discrete gauge group there must appear cosmic strings which are related to so-called `Alice strings' \cite{Schwarz:1982ec,Alford:1990mk,Bucher:1991qhl} and at low energies lead to a global two-form symmetry. It would be very useful to think further about the structure of such objects in particle physics contexts in light of generalized symmetry technology \cite{Heidenreich:2021xpr,McNamara:2022lrw,Arias-Tamargo:2022nlf}, as these two-form symmetries must be intertwined with the global zero-form and one-form symmetries as well. Furthermore, if the link fields condense at different scales as in recent deconstruction models, then domain walls may appear at intermediate scales, and it would be interesting to study their properties in detail. At low enough energies after all three fields have condensed and we only have the flavor-diagonal SM color, these domain walls eventually disappear as the $S_3$ has trivialized.

\subsection{Generating CKM}
 
\begin{figure}
    \centering
    \includegraphics[width=0.7\linewidth]{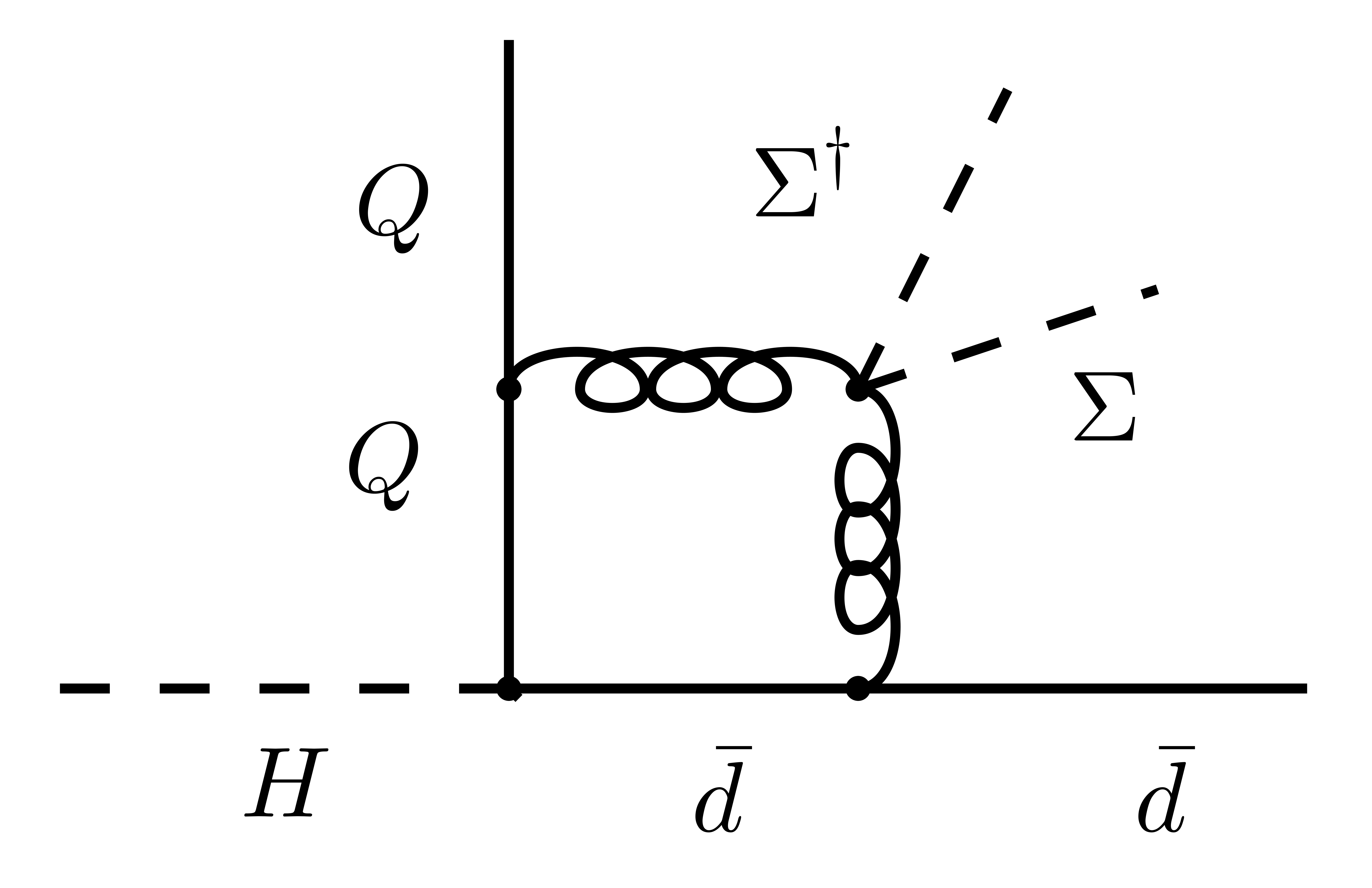}
    \caption{Flavor violation communicated to the down-type quarks through a loop of heavy gluons upon $SU(3)^3 \rightarrow SU(3)$. This yields complex, yet hermitian yukawas.}
    \label{fig:QLCFGluonLoop}
\end{figure}

\begin{figure}
    \centering
    \includegraphics[width=0.7\linewidth]{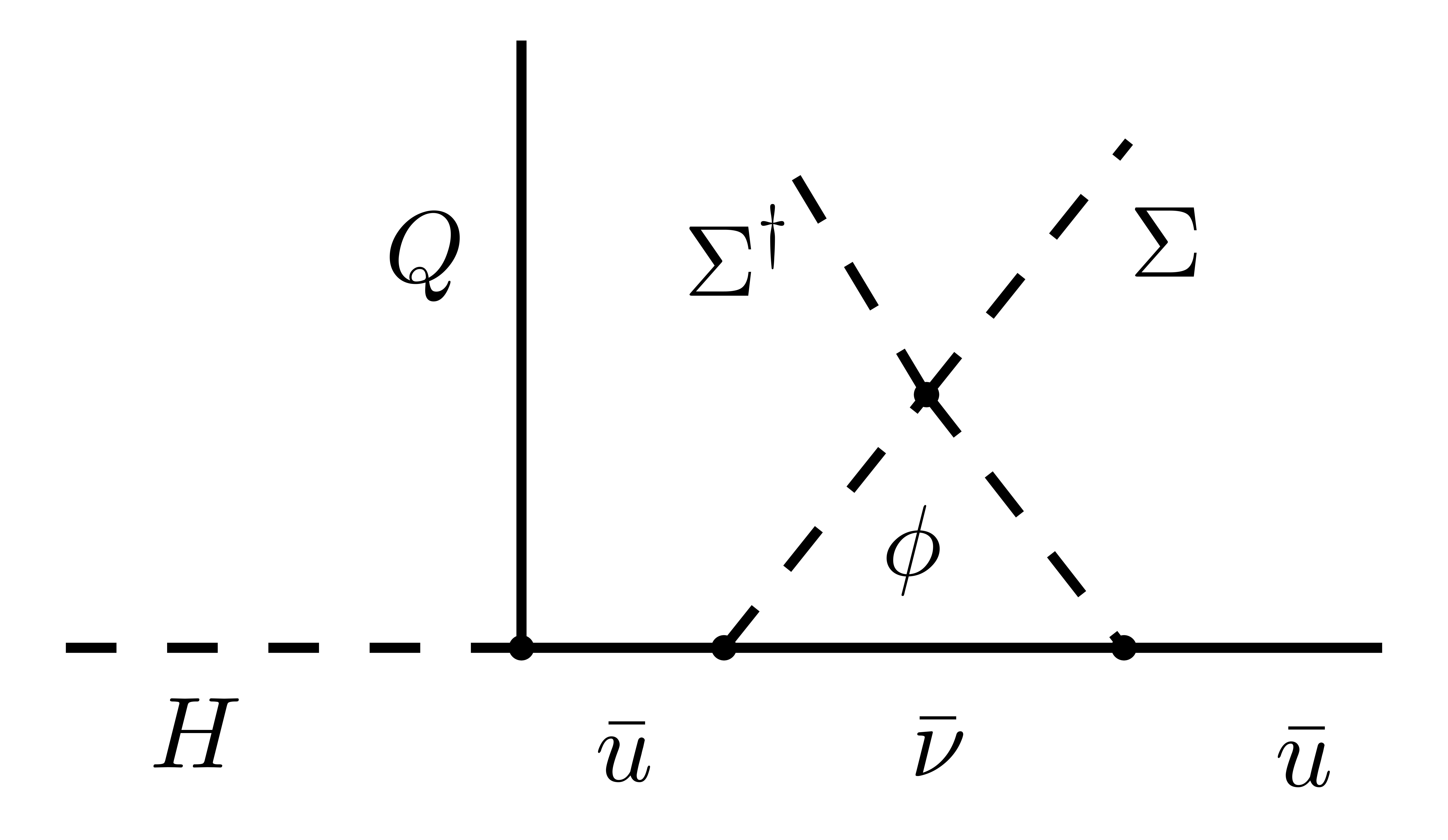}
    \caption{The up-type yukawas receive gluon loop corrections and additional loop-level flavor-violating corrections from $\phi^{a\alpha}$ and $\bar \nu$ interactions which are hermitian. The resulting yukawas still have real determinant.}
    \label{fig:QLCFScalarLoop}
\end{figure}

In \cite{Cordova:2024ypu} we showed that quark color-flavor $SU(9)$ unification could implement a flavor-symmetric massless quark solution to strong CP, and we refer there for more detail. In addition to generating the down quark yukawas from the up quark yukawas, a crucial component was the demonstration that general quark yukawas---and in particular a large CP-violating CKM angle---could be generated without upsetting the massless quark solution. The proof of principle scheme developed there worked as follows:
\begin{itemize}
    \item Fields arising from the $SU(9)$ complex adjoint $\Sigma^a_b$ spontaneously break the intermediate color-flavor symmetries down to $SU(3)_C$. These same fields also spontaneously or explicitly violate CP through the scalar potential, e.g. in the simplest case \begin{equation}
        V_{\slashed{\rm CP}}(\Sigma) \supset \eta_1 \text{Tr}(\Sigma^4) + \eta_2 \text{Tr}(\Sigma^2)^2 + \dots ,
    \end{equation}
    where we stress the mechanism works no matter how the CP-violation appears in the potential.
    \item The symmetries and spectrum (including in particular the absence of vector-like quarks) dictate that the breaking from $\Sigma$ is communicated to the quarks only at loop level, which generates complex yukawas that are automatically hermitian (in the basis where the tree-level yukawas are real, which always exists because of the UV PQ symmetry and the down-type yukawas having been generated by instantons). This results from the fact that for every diagram with an insertion of a $V_{\slashed{\rm CP}}$ interaction on the $\Sigma$ leg, there is a corresponding diagram with $V_{\slashed{\rm CP}}^\dagger$ inserted on the $\Sigma^\dagger$ leg. Then the yukawas generated have a structure 
    \begin{align}
    &(y_d)^a_{\ b} \sim \  y_b \left( \mathds{1}^a_{\ b} + \frac{\alpha}{4\pi} \frac{\lbrace \Sigma, \Sigma^\dagger \rbrace^a_{\ b}}{\Lambda_3^2} + \dots \right. \\ 
    & \left. + \frac{\alpha} {(4\pi)}\frac{\eta_1^\dagger (\Sigma^{\dagger 4})^a_{\ b} + \eta_2^\dagger \text{Tr}(\Sigma^{\dagger 2}) (\Sigma^{\dagger 2})^a_{\ b}}{\Lambda_3^4} + \dots + \text{ h.c.}\right), \nonumber
    \end{align}
    where the first line contains terms with manifestly real coefficients and the second line contains CP-violating terms with complex coefficients but which come always with a hermitian conjugate term. This means we can generate non-trivial $\delta_{\rm CKM}$ while manifestly preserving $\bar \theta = 0$. See Figure~\ref{fig:QLCFGluonLoop}.
    \item Loop communication through the heavy gluons alone would generate the same yukawa structure for the up and down yukawas, which would result in vanishing CKM phase, so another pair of fields must be present that distinguishes up and down quarks by communicating additional violation only to one type, again at loop level. See Figure~\ref{fig:QLCFScalarLoop}. 
\end{itemize}
In the $SU(9)$ theory, in order to find contributions distinguishing between up and down we needed to add new fields to the theory accomplishing this. However, here we already have $\phi^{a\alpha}$ and $\psi_{\ell u}$, which bear out the exact phenomenology needed. $\phi^{a\alpha}$ has a cubic or quartic coupling to $\Sigma^a_b$ and connects only the up-type quarks to the right-handed neutrinos via the $\lambda \phi^{a\alpha} \psi_{qu} \psi_{\ell u}$ interaction. This will generate contributions like
\begin{equation}\label{eqn:upYukawa}
    (y_u)^a_{\ b} \sim \frac{y_t}{y_b} (y_d)^a_{\ b} + y_t |\lambda|^2 c \frac{(\Sigma^\dagger \Sigma)^a_{\ b}}{M_\phi^2} + \dots + \text{ h.c.},
\end{equation}
where $\lambda$ is the coupling of $\phi$ to $\psi_u \psi_u$, $c$ is a quartic coupling, and these contributions skew $y_u$ and $y_d$ apart and allow the generation of a nontrivial $V_{\rm CKM}$. In the denominator $M_\phi$ is the mass of $\phi^{a\alpha}$, the component which participates in both quark and lepton flavor, which does not condense and may get a light mass. This structure appears for free simply from the embedding of $SU(9)$ color-flavor unification into $SU(12)$ unification along with our strategy to generate Majorana neutrino masses by breaking lepton flavor. 

While in previous work the intermediate group has been the flavor-unified $SU(3)^2/\mathbb{Z}_3$, here the intermediate group has undergone flavor deconstruction, but this does not affect the workings of this mechanism. In our case mixing will be generated when the bifundamental link fields $\Sigma_{ab}$ of Figure~\ref{fig:Quiver} spontaneously break $SU(3)^3 \rightarrow SU(3)_C$.\footnote{As an interesting connection we note Agrawal and Howe \cite{Agrawal:2017evu, Agrawal:2017ksf} considered a massless quark solution starting with $SU(3)^3_i$ and generating each generation's smallest yukawa from instantons of that generation. In this case they generate flavor-breaking using the link fields connected to the SM through dimension five operators e.g. $H Q_i \Sigma^i_j \bar u^j$. It would be interesting to determine if their model can be embedded in $SU(9)$ color-flavor unification through their suggested UV completion which adds vector-like quarks.}
The scalar potential in full generality contains many cubic and quartic terms, giving in principle enough freedom to generate the full CKM matrix by tuning the various coefficients to control the various entries of Eqn~\ref{eqn:upYukawa}. The development in this theory of a mechanism which would automatically generate a SM-like structure is an important goal for future work.

\section{Breaking Lepton Flavor to Hypercharge and Discrete Gauge Theory} \label{sec:breakingSU3}

Now we wish to break lepton flavor. We will condense not only the adjoint $\Sigma$, but also the two-index symmetric $\phi$. 
As in previous sections we will discuss the breaking in multiple stages, though these can occur at the same scale. 

Step 1: The complex adjoint $\Sigma^\alpha_\beta$ gets a vev which preserves solely the lepton-flavor direction $T^8$ = $U(1)_{L_i + L_j - 2L_k} \subset SU(3)_\ell$, of the form 
\begin{equation} 
\langle \Sigma^\alpha_\beta \rangle = \Lambda_\ell \begin{pmatrix}
        a_1 & c_1 & 0 \\ c_2 & a_2 & 0 \\ 0 & 0 & -(a_1 + a_2)
    \end{pmatrix}
\end{equation} where $a_i, c_i$ are complex and $c_1 \neq c_2^\star$. Under $SU(3) \rightarrow U(1)_{T^8}$ each real adjoint decomposes into four real singlets and two complex charge 3, and among the singlets there is a radial mode and the other $8-1=7$ modes are eaten by the heavy gauge bosons.

Step 2: The two-index symmetric $\phi^{\alpha\beta}$ gets a generic vev, which necessarily breaks the two continuous $U(1)$ directions down to hypercharge $Y =  (Q - N_cL) - 3R$. Under the lepton-flavored $U(1)_{T^8}$, the two-index symmetric has decomposed as $6 \rightarrow 2 (-1) + 3(+2) + 1(-4)$, written as multiplicity(charge). By writing the most general transformation of these under $U(1)_{T^8} \times U(1)_{Q-N_cL}$, one may find that hypercharge is not the only preserved symmetry, as under the linear combination $Q - N_cL - 6 T^8$, the two-index symmetric components have charges $2(0) + 3(-18) + 1(+18)$. As a result the condensation preserves 
\begin{equation}
    \mathbb{Z}_{18} \subset U(1)_{Q-N_cL - 6(L_i + L_j - 2L_k)},
\end{equation}
which we hereon refer to as $\mathbb{Z}_{18}^X \subset  U(1)_X$ for convenience.

Furthermore, the global structure appearing after this breaking is 
\begin{equation} \label{eqn:IRglobStruc}
    \frac{G_{\rm color} \times U(1)_Y \times \mathbb{Z}_{18}^X}{\mathbb{Z}_3 \times \mathbb{Z}_3},
\end{equation}
with both possible diagonal $\mathbb{Z}_3$ subgroups being set equal to the identity. The $\mathbb{Z}_3$ quotient between color and hypercharge arises out of the quotient between $Q-N_cL$ and the color group, which above we saw appeared at the first step of breaking. Then since $Y = (Q-N_cL) - 3R$, we have $Y$ (mod 3) = $Q-N_cL$ (mod 3), meaning the $\mathbb{Z}_3$ subgroup is blind to the difference and hypercharge inherits this overlap. The same too with $X$.

\subsection{Proton stability and tri-nucleon decay}
What are the consequences of this discrete symmetry? We may rewrite it as 
\begin{equation}
    Q - 9 (L_i + L_j - L_k) = 3\left[B - 3(L_i + L_j - L_k)\right],
\end{equation}
where the right-hand side, divided through by three, is the faithfully-acting symmetry after confinement.
As noted in \cite{Babu:2003qh,Davighi:2022qgb}, this has the remarkable consequence of forbidding proton decay! Under this infrared symmetry the baryons have charge $+1$ whereas the leptons have $\pm 3$ depending on their generation. Thus the proton is the lightest particle with nonzero charge $(\rm{mod} \ 3)$, and is exactly stable. We recall that already in the Standard Model the proton is stable as a result of an exact, anomaly-free $\mathbb{Z}_{2N_g}^{B+L}$ global symmetry in the infrared, in addition to the fact that we have $N_g>1$ generations. We see that in quark-lepton color-flavor unification, this same fact of proton stability being guaranteed by the existence of multiple generations is upgraded into a gauge symmetry which is now respected even by quantum gravity.\footnote{Note that such a discrete symmetry stabilizing the proton also requires only $N_g > 1$ in the generalization of this model to any $SU(N_g(N_c+1))$ quark-lepton color-flavor unification. That is, while the non-invertible symmetry effects in the quark sector work specially because $N_g = N_c$ and in the lepton sector work specially because $N_g > 2$, the ability of this model to stabilize the proton with a discrete gauge symmetry does not provide further insight into the generation puzzle.}

The baryon-number violating interactions consistent with this symmetry change baryon number by three, for example $3 \mathcal{N} \rightarrow \bar \ell_i + \bar \ell_j + \bar \ell_k$, where $\mathcal{N}$ is a nucleon and $\ell_i$ is an appropriate charged or neutral antilepton of the $i$th flavor. Inclusive searches for tri-nucleon decays have been performed, for example in germanium \cite{GERDA:2023uuw}, with claimed lower bounds on the partial life-times of order $10^{26}$ years. In our model the rate is extremely suppressed to the point of unobserveability, as this process comes from a dimension 18 operator when written in terms of quarks and leptons, and our symmetry-breaking dynamics occurs at quite high scales. Still it would be interesting to better understand the constraints on all possible tri-nucleon decay channels, and the extent to which this can teach us about low-energy BSM physics which is consistent with other constraints \cite{Koren:2026c}. 

\subsection{Lepton masses and mixings}

The loop-induced mechanism for quark flavor breaking works in the same way in the lepton sector, though here the leptonic gluons transfer flavor-breaking effects from both $\Sigma$ and $\phi$ to the lepton yukawas, and we get a schematic structure
\begin{align}
    (y_e)^\alpha_{\ \beta} &\sim \  y_\tau \left( \mathds{1}^\alpha_{\ \beta} + \frac{\alpha}{4\pi} \frac{\lbrace \Sigma, \Sigma^\dagger \rbrace^\alpha_{\ \beta} + (\phi^\dagger \phi)^\alpha_{\ \beta}}{\Lambda_\ell^2} + \dots \right. \\ 
    &\left. + \frac{\alpha} {(4\pi)} \frac{\eta^\dagger (\Sigma^{\dagger 4})^\alpha_{\ \beta} + a^\dagger (\phi^\dagger \Sigma^2 \phi)^\alpha_{\ \beta} + \dots }{\Lambda_\ell^4}  + \dots + \text{ h.c.}\right). \nonumber
    \end{align}
Here again the presence of CP violating couplings in the scalar potential allows CP-violation to transfer to the yukawas, and we emphasize there are many terms in the $V(\Sigma,\phi)$ potential which can all contribute to the generated yukawas.   

However, $\phi$ also generates Majorana masses for the right-handed neutrinos, which communicates additional flavor breaking through the flavorful type-I seesaw mechanism. In concert with the large $y_{\rm t}$-sized Dirac neutrino yukawas, this generates Majorana masses for the active neutrinos. Since the leptons all have charge $\pm 9$ under the discrete $\mathbb{Z}_{18}^X$ symmetry, the right-handed Majorana mass matrix is completely allowed. This comes from the structure
\begin{equation}
    \mathcal{L} \supset \half \lambda \phi^{\alpha \beta} \bar \nu_\alpha \bar \nu_\beta,
\end{equation}
and we see that if all the components of $\phi^{\alpha\beta}$ condense then we fully populate the Majorana neutrino mass matrix $M^{\alpha \beta} = \lambda \langle \phi \rangle^{\alpha\beta}$.

\begin{figure}
    \centering
    \includegraphics[width=0.7\linewidth]{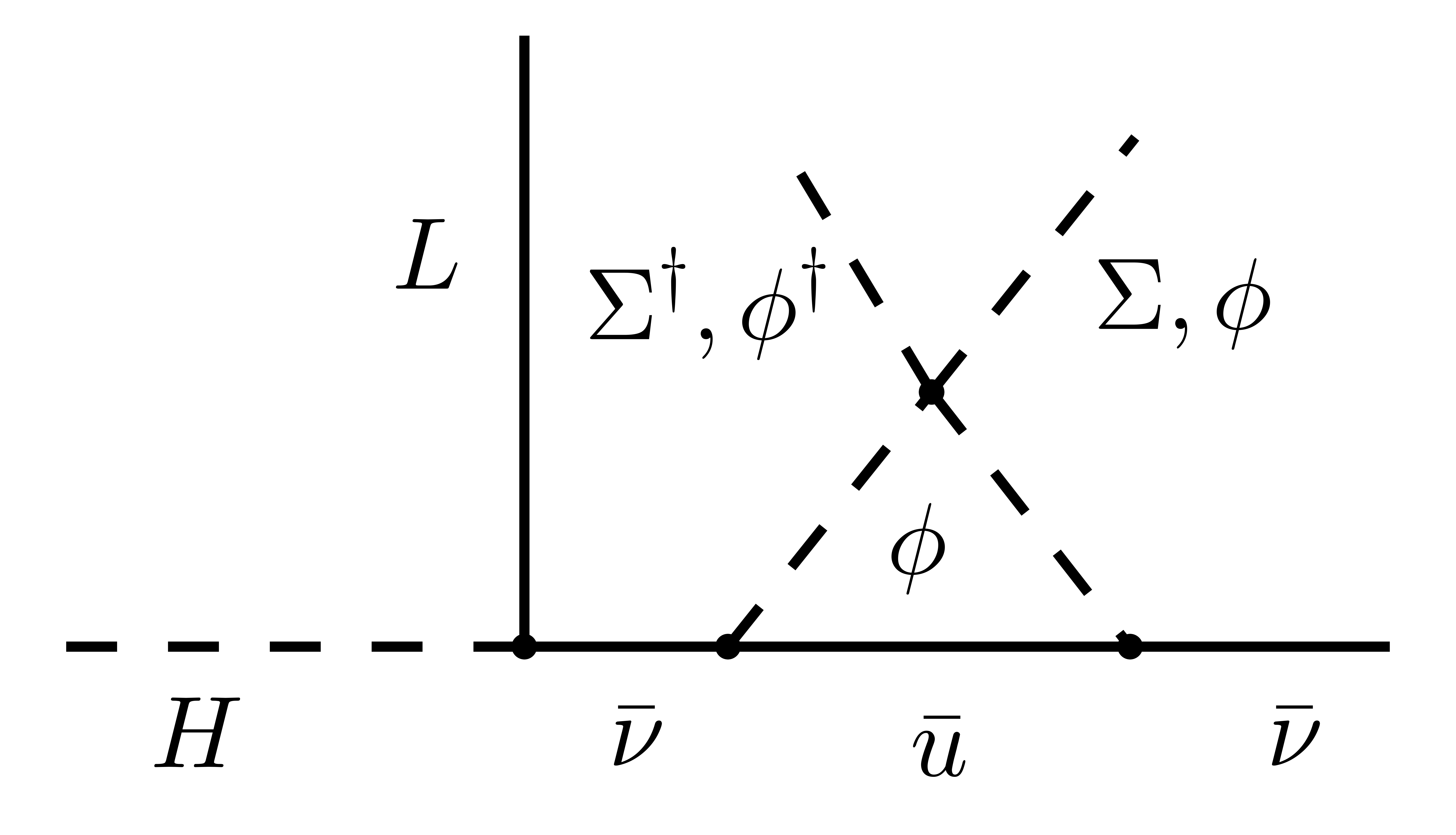}
    \caption{The Dirac neutrino yukawas receive leptonic gluon loop corrections and additional loop-level flavor-violating corrections from $\phi^{a\alpha}$ and $\bar u$ interactions.}
    \label{fig:QLCFScalarLoop2}
\end{figure}

Together these implement the type-I seesaw. Taking a general Dirac yukawa $y^\alpha_{\nu\ \beta}$ (since $y_t \delta^\alpha_{\ \beta}$ also gets corrections as in Figure~\ref{fig:QLCFScalarLoop2}), the resulting active neutrino masses are 
\begin{equation}
    {m_{\nu}}_{\alpha\beta} = v^2 {(y_\nu)}_{\ \alpha}^{\gamma} (M^{-1})_{\gamma \delta} {(y_\nu)}^{\delta}_{\ \beta},
\end{equation} 
where $(M^{-1})_{\gamma \delta} M^{\delta \sigma} = \delta_{\ \gamma}^{\sigma}$.
This seesaw effect can skew the PMNS matrix into anarchy even if the lepton Dirac yukawas have a similar structure to the quarks, and we have enough freedom in the $\Sigma, \phi$ vevs to accommodate arbitrary masses and mixings. The other fact we must match is a large `Dirac phase', which is the only one we are sensitive to observationally as yet, and arises from 
\begin{equation}
    \delta_{\rm PMNS} \propto \text{Im} \det \left[m_e m_e^\dagger, m_\nu m_\nu^\dagger\right]
\end{equation}
where $m_e = y_e v$, having a parallel structure to the CKM phase. With both $\Sigma, \phi$ the general potential will have both of their vevs complex, providing additional sources of CP violation compared to the quarks, meaning there is no problem attaining large CP violation in the neutrino sector. 

\section{Matching SM Gauge Couplings} \label{sec:matching}

Here we will consider the one-loop evolution of the gauge couplings of this model all together to ensure we can match on to the SM gauge couplings at electroweak scales. In this section alone we will consider the conventional normalizations for the gauge couplings, such that e.g. the right-handed down quark has $Y = 1/3, \ B = -1/3, \ R = -1/2$.

We will consider the simplest case where all around the the Majorana mass scale we have 
\begin{equation}
    SU(12) \times U(1)_R \rightarrow \frac{G_{\rm color} \times U(1)_Y \times \mathbb{Z}_{18}^X}{\mathbb{Z}_3^2},
\end{equation}
and at this stage the color group is $(SU(3)^3\times U(1)^2)/\mathbb{Z}_3^2$. That is, $SU(12)$-breaking, $SU(9)$ quark color-flavor breaking, and $SU(3)$ lepton flavor breaking happen nearly simultaneously at the scale $\Lambda_{12}$, as does the breaking of $U(1)_R \times U(1)_{B-N_cL} \rightarrow U(1)_Y$. We also consider $G_{\rm color} \rightarrow SU(3)_C$ all at once at a lower scale $\Lambda_3$. It's possible for $\Lambda_3$ to also be quite high, but a low $\Lambda_3$ has the benefit that flavor-changing effects may be observable in the near-term. 

Our model has not added any $SU(2)_L$-charged matter, so $g_L$ evolves as in the SM at all energies. Below the scale $\Lambda_{12}$ there are also no new states charged under hypercharge, so the hypercharge coupling also evolves as in the SM up to this scale. In the usual normalizations, we have $Y = (B-L)/2 - R$, so at this scale the hypercharge coupling matches to the $R, Q-NcL$ couplings as 
\begin{equation}
    \alpha_Y^{-1} = \alpha_R^{-1} + \frac{1}{4}\alpha_{B-L}^{-1},
\end{equation}
and at this scale $\alpha_{B-L} = \alpha_{12}/8$. Then we can determine the $\alpha_R$ needed as 
\begin{equation}
    \alpha_R^{-1} = \alpha_Y^{-1} - 2 \alpha_{12}^{-1}.
\end{equation}
Since $\alpha_R$ is a free parameter, this works as long as $\alpha_Y^{-1} > 2 \alpha_{12}^{-1}$ at this scale, which will easily be the case. The size of $\alpha_{12}$ is set by the requirement of generating the correct down-type yukawas---this can only be accurately calculated on the lattice, so we simply take $\alpha_{12}(\Lambda_{12})=1$ as a benchmark.\footnote{While this is large, recall that the power-counting for perturbative corrections is in $\alpha/(4\pi)$, so our treatment is still sensible.}

At the electroweak scale we measure (see the PDG \cite{ParticleDataGroup:2024cfk} for more precision)
\begin{equation}
    \alpha_Y^{-1} \simeq 98, \quad \alpha_L^{-1} \simeq 30, \quad \alpha_C \simeq 8.5,
\end{equation}
where we note we are of course not using the modified hypercharge gauge coupling by a factor $3/5$ as is done in studies of unification into $SU(5)$ or $SO(10)$. 

\begin{figure}
    \centering
    \includegraphics[width=1\linewidth]{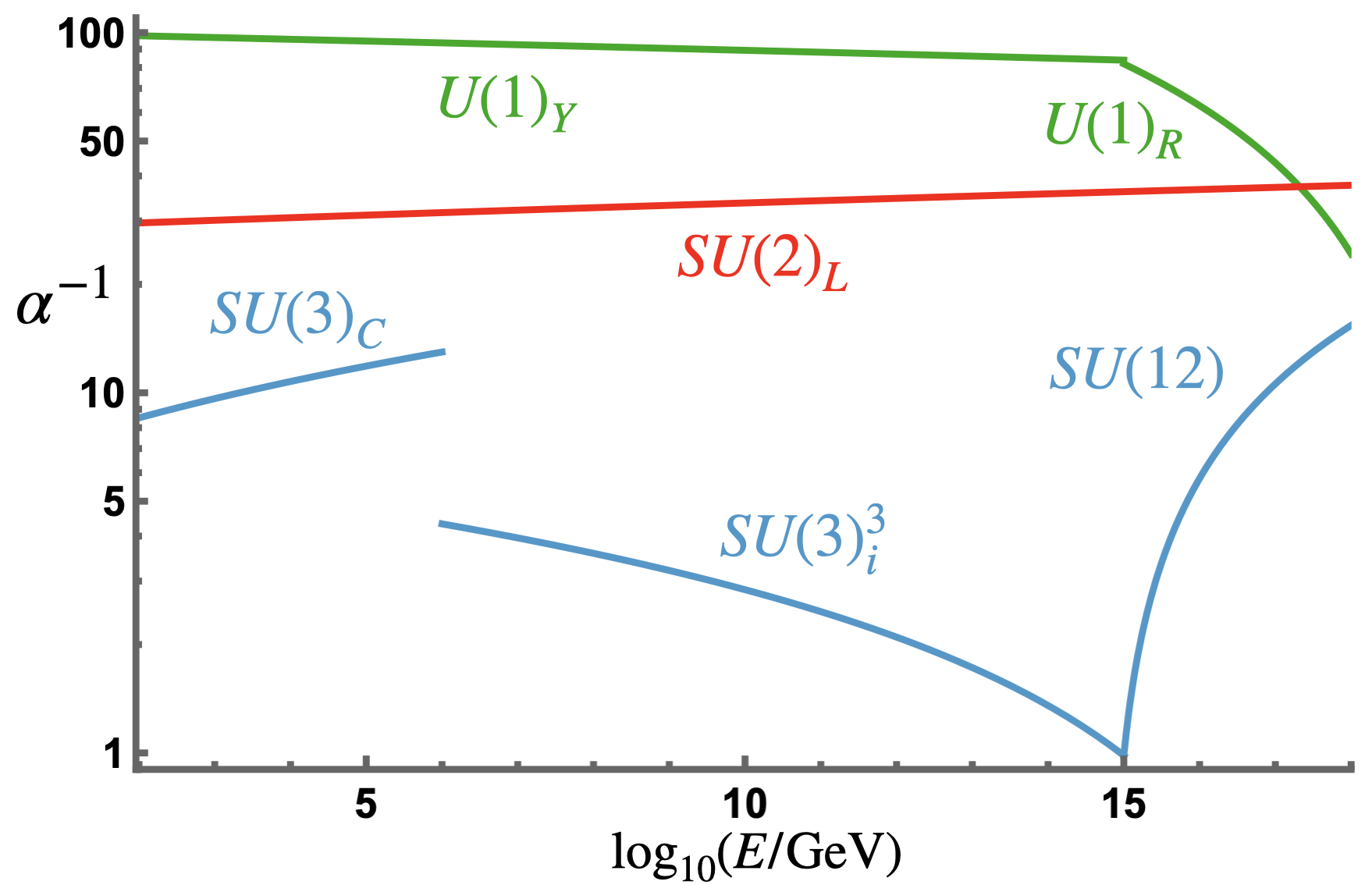}
    \caption{The couplings as a function of scale in the simple benchmark model where most of the breaking steps occur simultaneously at $\Lambda_{12}=10^{15}$ GeV. Non-trivial matching occurs in $U(1)_Y$ emerging out of $U(1)_R$ and $SU(12)$, and at $SU(3)^3 \rightarrow SU(3)$ breaking at $\Lambda_3 = 10^6$ GeV.}
    \label{fig:RunningCouplings}
\end{figure}

In Figure~\ref{fig:RunningCouplings} we plot an example case with $\Lambda_{12} = 10^{15}$ GeV, such that our seesaw mechanism works successfully, and $\Lambda_3 = 10^6$ GeV, such that we avoid constraints on FCNCs. As mentioned above, we require some colored spectator fields and we take $\beta_{\rm spectator} = 37/6$ to help turn around the $SU(3)_i$ gauge couplings so they are no longer asymptotically free between these two scales, and we assume these extra degrees of freedom decouple at $\Lambda_3$. A different number of these would simply modify $\alpha_{12}(\Lambda_{12})$.

Since the ultraviolet group is not simple, no gauge coupling unification is demanded, but it is interesting if it is achievable nonetheless as it points to the possibility of further ultraviolet unification. In this example we see that the $U(1)_R$ and $SU(2)_L$ couplings meet up around $10^{17}$ GeV, and their unification would be consistent with the ultraviolet completion mentioned in the introduction of $(SU(2)_R \times SU(2)_L )\rtimes \mathbb{Z}_2$.  To further unify with the $SU(12)$ coupling would require additional matter with electroweak quantum numbers. One possibility, as mentioned in the introduction, is to upgrade the Higgs sector to $SU(12)$ adjoint+singlet fields, and perhaps to include $H_u$ and $H_d$ separately. These fields could help with the running, as well as having useful roles to play in generating the quark masses and mixings with less tuning required. We leave for future work the possibility of gauge coupling unification in models like this. 

\section{One- and Two-Form Symmetries of the Standard Model} \label{sec:SMsyms}

An interesting feature of the infrared of this theory is that we do not land on the familiar Standard Model \emph{group}. Recently attention has been called to the fact that the SM has not been uniquely defined by data or theory. The so-far explored family of field theories differ in the `global structure' of the gauge group, or in gauge-invariant language in the one-form global symmetry of the theory \cite{Tong:2017oea, Reece:2023iqn,Choi:2023pdp,Cordova:2023her,Koren:2024xof}. However, this is far from the only family of field theories which have the same local-in-field-space phenomenology of the familiar SM, and the studied possibilities do not even exhaust the possible global structures of the SM. 

In particular, there are interesting versions of the SM which have additional discrete gauge theory factors. 
In gauge-invariant language, this leads to two-form global symmetries of the Standard Model. These factors may appear simply out of well-motivated UV completions, as we have seen from the dynamics above which have led to the SM group in this theory being

\begin{equation}
    G_{\rm SM} = \frac{SU(3)_C \times SU(2)_L \times U(1)_Y \times \mathbb{Z}^X_{18}}{\mathbb{Z}_3 \times \mathbb{Z}_3 \times \Gamma},
\end{equation}

\noindent with $\Gamma \in \lbrace 1,\mathbb{Z}_2\rbrace$ the global structure we began with in the ultraviolet which has descended to the electroweak global structure.

If we were to instead study the Standard Model with $\mathbb{Z}^X_{18}$ gauged and no quotient between $X$ and $Y$, that theory would have a $\mathbb{Z}_3^{(1)}$ electric one-form symmetry. This would correspond to the absence of particles with $Y \neq X \ (\text{mod } 3)$, and resultingly that Wilson lines in such representations, despite being allowed in this theory, are not `endable'. The theory under consideration here gauges that $\mathbb{Z}_3^{(1)}$ symmetry (discussed further below) and this forbids such particles from existing, which makes searching for such `fractionally $X$-onic' particles a possible low-energy probe of this model.\footnote{In spirit this is just the same as how fractionally electrically charged particles probe the usual Standard Model structure, although here it is much more difficult to test at a collider as it requires determining the $X$ charge of a discovered particle. Since we have no massless $X$ vector, this is not generically possible, but one can write down models where discovering new fields and measuring multiple couplings would allow one to reconstruct the $X$ charge. If a particle with $Y \neq X \ (\text{mod } 3)$ is discovered, this model is flatly ruled out.} 

To our knowledge this is the first time a particle physics model has been identified that contains a quotient shared between continuous and discrete gauge groups. 
The systematic construction of Standard Models with two-form symmetries, and their interplay with the possible one-form symmetries, is being considered in ongoing work. In the rest of this section we just make a few further comments on the expected physics, saving detailed analysis to be presented in the near future.

There have only been few investigations of versions of the Standard Model which have discrete gauge symmetries.\footnote{Discrete gauge symmetries have been considered more often in the context of the MSSM, starting from \cite{Bento:1987mu,Ibanez:1991hv,Ibanez:1991pr}.} The most studied possibility, once right-handed neutrinos are included, is a gauged $\mathbb{Z}_N$ discrete subgroup of $B - L$. The case of $N=2$ appears, for example, from $SO(10)$ models bearing out the type 1 seesaw, with the cosmic strings first discussed in \cite{Jeannerot:1996yi}. For larger $N$ to our knowledge the only works considering the cosmic strings are our earlier \cite{Craig:2018yvw,Koren:2022axd}. 

When this arises from a gauged $U(1)_{B-L}$ which is spontaneously broken, for general $N$ this bears out a charge-$N$ Abelian Higgs theory and at low energies one arrives at a `BF theory', which describes the universal infrared dynamics of pure $\mathbb{Z}_N$ discrete gauge theory.\footnote{See e.g.  \cite{Brennan:2023mmt,Banks:2010zn} for derivation and further discussion.} The fields consist of a one-form gauge field $A^X_\mu$ and a two-form gauge field $B_{\mu\nu}$ which is the dual of the pseudoscalar phase of the Higgs $\phi$, the continuous part of which has become the longitudinal mode of $A^X$. They interact via the Lagrangian
\begin{equation}\label{eqn:BFtheory}
    \mathcal{L} \supset \frac{N}{2\pi} \partial_\mu A^X_\nu B_{\rho \sigma} \epsilon^{\mu\nu\rho\sigma} + J^\mu A^X_\mu,
\end{equation}
where $J^\mu$ is the $X$ current inherited from the theory with continuous gauge symmetry.  
In the farthest IR, having integrated out the charged particles and the cosmic strings, this theory has emergent one- and two-form global symmetries $\mathbb{Z}_N^{(1)} \times \mathbb{Z}_N^{(2)}$. The charged objects are Wilson lines $\exp i q \oint A^X$ (electric particle worldlines) and Wilson surfaces $\exp i k \oint B$ (cosmic string worldsheets) which interact only through topological linking. 

In our theory, the condensation of $\phi^{\alpha \beta}$ in the early universe produces cosmic strings around which both $\phi^{\alpha \beta}$ and the $U(1)_X$ gauge field $A^X$ wind. These strings all have magnetic fluxes running through their cores. Those strings with integer magnetic fluxes in units of the Dirac flux of $U(1)_{Q-N_cL}$ and $U(1)_{L_i + L_j - 2 L_k}$ are unstable to the pair-production of magnetic monopoles. As a result of the non-trivial $N=18$ charge of $\phi$, there are additional \textit{stable} cosmic strings which have fractional magnetic fluxes which cannot be discharged by monopole-antimonopole production, often called `BF strings'. 

However, in relation to the general charge-$N$ Abelian Higgs our theory with $X = Q- 9(L_i + L_k - L_k)$ becomes more complicated in two different ways, both of which mean that the BF theory is only schematic in this case. The first complication is that the condensation of $\phi^{\alpha \beta}$ not only breaks $U(1)_X$ but also produces a large Majorana mass for right-handed neutrinos. This issue also arises in models of gauged $U(1)_{B-L}$ spontaneously broken to $\mathbb{Z}_2$ \cite{Sahu:2004ir,Jeannerot:1996yi,Gu:2005gu}. The large Majorana mass appears everywhere apart from the cores of cosmic strings, where $\langle \phi^{\alpha \beta} \rangle = 0$ is restored and the right-handed neutrino modes are massless. Furthermore, the topological structure of the Dirac operators in these twisted gauge field backgrounds necessitates the existence of zero-energy solutions \cite{Callan:1984sa,Witten:1984eb}. Together these effects lead to localized $\bar \nu$ zero modes on strings. Technically, this extra coupling complicates the dualization of the longitudinal mode of $\phi$ to $B_{\mu\nu}$. Physically, the fermi zero-modes change the physics of the cosmic strings, making them superconduct lepton flavor currents, and the far infrared theory must be adjusted in some manner to account for this. To our knowledge understanding the correct far-infrared description remains an important open question. 

Furthermore, in the case of a BF theory the two-form symmetry is connected to the SM only through the one-form gauge field's coupling to a SM current, which physically leads to the effect of the SM particles picking up Aharonov-Bohm phases around the cosmic strings, and in this case leptons of different flavors could pick up opposite phases around the string. Such flavored cosmic strings have received only little attention \cite{Chkareuli:1991tp,Bibilashvili:1990qm}.
However as we've seen in the example here, there can also be a subtler connection between the new two-form symmetry and the dynamics of the SM. In this case the zero-form gauge group has a quotient between subgroups of color, hypercharge, and the discrete $X$ group, which means that the one-form symmetry structure affects the two-form symmetry structure by correlating the magnetic fluxes. 

For now we ignore the former issue regarding zero-modes. Then we can construct the $(U(1)_Y \times \mathbb{Z}_{18}^X)/\mathbb{Z}_3$ theory by starting with the $(U(1)_Y \times \mathbb{Z}_{18}^X)$ theory and implementing the quotient by gauging the $\mathbb{Z}_3^{(1)}$ electric one-form symmetry which acts on both the hypercharge and the discrete gauge sectors as a shift of the gauge fields $A^X_\mu \rightarrow A^X_\mu + \Lambda_\mu, \ A^Y_\mu \rightarrow A^Y_\mu + \Lambda_\mu$ with $\partial_{[\nu} \Lambda_{\mu]} = 0$. This construction can be thought of in two steps: First, we couple the theory to a background two-form gauge field $C_{\mu\nu}$ transforming as $C_{\mu\nu} \rightarrow C_{\mu\nu} + \partial_\mu \Lambda_\nu$ in such a way to make the theory formally invariant under arbitrary one-form transformations $\partial_{[\nu} \Lambda_{\mu]} \neq 0$. Then we gauge this one-form symmetry by summing over $C_{\mu\nu}$ in the path integral, which requires including discrete dynamics for $C_{\mu\nu}$ by giving it a one-derivative kinetic term in the form of a BF theory with a new one-form field $E_\mu$,
\begin{align} \label{eqn:IRBF}
    \mathcal{L} = &\frac{1}{4g_Y^2}\left(\partial_{[\mu} A^Y_{\nu]} - C_{\mu\nu} \right)^2 + \frac{18}{2\pi} \left(\partial_\mu A^X_\nu  - C_{\mu\nu} \right)B_{\rho \sigma} \epsilon^{\mu\nu\rho\sigma} \nonumber \\
    & + \frac{3}{2\pi} \partial_\mu E_\nu C_{\rho \sigma} \epsilon^{\mu\nu\rho\sigma}.
\end{align}
Here the first term is the familiar kinetic term for hypercharge coupled to the new two-form gauge field $C_{\mu\nu}$ so it is invariant under the gauged one-form symmetry, the second term is the discrete kinetic term (BF term) for $A^X_\nu$ after having been made invariant under the gauged one-form symmetry and which enforces that $A^X_\nu$ gauges a $\mathbb{Z}_{18}^{(0)} \subset U(1)^{(0)}$ symmetry, and the final term is a BF term for $C_{\rho\sigma}$ itself which enforces that it gauges a $\mathbb{Z}_{3}^{(1)} \subset U(1)^{(1)}$ symmetry. As appropriate, the BF terms also enforce that the discrete gauge fields are restricted to be flat (have vanishing field strength) as a discrete field has no propagating degrees of freedom. 

We will return in future work to study this low-energy theory in further detail, along with more generally the possible two-form symmetries of the Standard Model. One physical takeaway from this section is that the cosmic strings produced in this theory interact with the SM degrees of freedom not only through Aharonov-Bohm scattering, but also in a subtler way described (as yet obscurely) by Equation~\ref{eqn:IRBF}, and it would be very interesting to understand their effects on, for example, the cosmic neutrino background. The applicable constraints on these cosmic strings depend sensitively on their interactions with the plasma and their emission spectra, and require further study. A broader conceptual point that is exemplified here is that there are various interesting ways to couple a topological quantum field theory to the SM,\footnote{Note that in the second and third terms spacetime indices are contracted solely with the Levi-Civita psuedotensor, so there is manifestly no metric dependence in these TQFT factors. Note also that Juven Wang and collaborators have compiled much topological data on the SM and GUTs \cite{Wan:2019gqr,Wang:2020xyo,Wang:2020gqr,Wang:2020mra,Wang:2021hob,Wang:2021ayd,Wang:2021vki,Wang:2022eag,Putrov:2023jqi}.} and the phenomenology of these possibilities has hardly been explored. 

\section{Conclusion} \label{sec:conclusion}

In this work we have proposed quark-lepton color-flavor unification as a novel vista in the landscape of ultraviolet theories where microphysics gets simpler at small distances, providing a radical conceptual departure from conventional vertical theories of unification. The generation structure of the Standard Model is non-trivially incorporated into the familiar forces simply by gauging some of the approximate global symmetries of the SM fermions, providing multiple structural benefits. The SM matter is unified into few representations without requiring new fermions. The starkly disparate phenomenologies of quarks and leptons nonetheless emerge from a unified, highly symmetric ultraviolet origin. Most strikingly the proton is absolutely stable as the result of a discrete gauge symmetry that emerges naturally in the infrared, which to our knowledge is the only possibility for absolute proton stability in the landscape of four-dimensional unified theories without additional fermions.

Here we have established this framework and exhibited interesting phenomenologies arising from both perturbative and nonperturbative aspects of this gauge theory, as it breaks down from the ultraviolet theory to the Standard Model phase. In the ultraviolet we started with a single yukawa coupling shared between the up-type quarks and the neutrinos. In the quark sector, instantons generated the bottom yukawa and resolved the strong CP problem in a flavor-symmetric massless quark solution. In the lepton sector, instantons break a noninvertible symmetry to generate the tau yukawa and explain the hierarchy between the heaviest quarks and the heaviest leptons, which also sets up a flavored type-I seesaw mechanism that flips the charged vs. neutral mass hierarchy. The symmetries have special structure because $N_g = 3$ specifically, providing one of few potential inroads toward the generation puzzle.

Intense study of the full scope of phenomenology in this framework is surely merited, and we have only begun to explore the structure. An obvious and important target is a better handle on the scalar sector, including how easily achievable the required vevs are, and how to generate a realistic flavor structure without requiring fine-tuning. As to the extra hierarchy problems, we note that the usual solutions (which seem not be realized at the weak scale for the electroweak hierarchy problem) still work fine to stabilize these flavor scales, and it would be interesting to study e.g. a supersymmetric version of this theory. We have suggested that upgrading the SM Higgs to a 2HDM with color-flavor adjoints + singlets may be a useful direction here. Getting a handle on these aspects will reveal the most testable aspects of this theory in precision experiments and possibly at colliders, for example the related \cite{Davighi:2022qgb,Greljo:2022dwn}, though many more near-term signatures would be available in a variant of this theory that lowered the scale of lepton flavor symmetry-breaking.

There is also much interesting physics in the cosmology of this theory, where the various defects we have seen appear at intermediate scales will be brought to life, especially if the Higgsing steps are more spread out than in our benchmark model. As the flavor gauge symmetries are broken, a wide variety of magnetic monopoles appear: flavor-universal $Q-N_cL$ monopoles, both with and without non-Abelian color-flavor magnetic fluxes; flavored $Q_i - Q_j$ quark family difference monopoles, which may also have flavor-specific color fluxes; and lepton family difference monopoles, all of which will annihilate automatically at later stages when the $U(1)$ symmetries are broken and the monopoles become the endpoints of cosmic strings. Alice strings appear as the result of a gauged discrete semi-direct product group, and are later destroyed when they become the boundaries of domain walls, which themselves disappear upon infrared color reunification. Still other strings are produced when the infrared discrete $X$ symmetry emerges, and these both superconduct lepton flavor currents and are stabilized by fractional magnetic fluxes. These dynamical topological defects deserve dedicated further study, and may have interesting nonperturbative effects on the SM plasma in the early universe. They will also produce a distinct spectrum of gravitational waves between the appearances, interactions, and disappearances of these varied types of nonperturbative objects, which will provide unique, testable signatures of this model in future gravitational wave observatories.
Further fruitful topics in the cosmology of this model include identifying ways in which additional required BSM physics is naturally incorporated into this framework. The potential mechanisms for baryogenesis are manifold, between leptogenesis \cite{Greljo:2025suh}, the possibility of first-order flavor-breaking phase transitions and bubble wall dynamics, flavored topological defects, and the possible inclusion of axions \cite{Berbig:2026a}. This model also provides a top-down framework for the existence of dark matter candidates with SM flavor quantum numbers.

Finally, the infrared appearance of an additional discrete structure coupled to the Standard Model degrees of freedom further heralds a new program of studying phenomenological implications of topological quantum field theory factors. Much of the recent work on generalized symmetries in particle physics motivates that such topological degrees of freedom can have important phenomenological effects. Indeed the many dynamical topological defects which appear at high energies in this model can leave behind important, subtle effects in the infrared, as we have seen with the non-invertible chiral symmetries which depend on the structure of magnetic one-form symmetries. Understanding more fully the generalized symmetry structure at each phase of this theory may reveal additional striking phenomenology. 

By considering a maximalist interpretation of the experimental data, we have found a novel alternative to the types of unification on which decades of study has been spent. In this framework the surprising stability of the proton does not portend epicycles on old theories, but is rather a harbinger of the deep intertwining of the Standard Model with topological degrees of freedom. 

\section*{Acknowledgments}

We thank Adam Martin and Fengwei Yang for helpful conversations and Maximilian Berbig and Adam Martin for comments on a draft of this manuscript. SK is grateful for inspiration from Gustavo Corti\~{n}as, as well as Stephen \& Mary Merriman and all the players at the  Merrimans' Playhouse jazz open sessions in South Bend. We have made use of FeynGame \cite{Bundgen:2025utt} to create figures, and throughout this work we have found LieART \cite{Feger:2019tvk} to be a helpful resource. This work is partially supported by the National Science Foundation under grant PHY-2412701.

\let\oldaddcontentsline\addcontentsline
\renewcommand{\addcontentsline}[3]{}
\bibliography{qlcfu}
\let\addcontentsline\oldaddcontentsline

\end{document}